%% file: main.tex
\documentclass[12pt]{article}

\usepackage[letterpaper, margin=1in]{geometry}
\usepackage[utf8]{inputenc}
\usepackage{times}
\usepackage{graphicx}
\usepackage{mathptmx}
\usepackage{amsmath}
\usepackage{dsfont}
\usepackage{amsfonts}
\usepackage[justify]{ragged2e}
\usepackage[labelfont=bf,labelsep=period]{caption}
\usepackage{lineno}
\usepackage[style=nature,maxbibnames=20,url=false,eprint=false,isbn=false]{biblatex}
\usepackage{hyperref}
\usepackage[section]{placeins}

\DeclareUnicodeCharacter{2212}{-}

\hypersetup{
    colorlinks=true,
    linkcolor=blue,
    filecolor=magenta,      
    urlcolor=cyan,
}
\usepackage{titling}

\title{First Order Quantum Phase Transition in the Hybrid Metal-Mott Insulator Transition Metal Dichalcogenide 4Hb-TaS$_2$}
\author{Abhay Kumar Nayak$^{1}$, Aviram Steinbok$^{1}$, Yotam Roet$^{1}$,\\ Jayhun Koo$^{1}$, Irena Feldman$^{2}$, Avior Almoalem$^{2}$, Amit Kanigel$^{2}$,\\ Binghai Yan$^{1}$, Achim Rosch$^3$, Nurit Avraham$^{1\dagger}$, Haim Beidenkopf$^{1\dagger}$}
\date{\small $^{1}$ Department of Condensed Matter Physics, Weizmann Institute of Science, Rehovot, Israel.\\
$^{2}$ Department of Physics, Technion - Israel Institute of Technology, Haifa 32000, Israel.\\
$^3$ Institute for Theoretical Physics, University of Cologne, 50937 Cologne, Germany.\\
$^{\dagger}$ Corresponding authors: nurit.avraham@weizmann.ac.il, haim.beidenkopf@weizmann.ac.il}

\begin{document}

\maketitle

\section*{Abstract}
\textbf{Coupling together distinct correlated and topologically non-trivial electronic phases of matter can potentially induce novel electronic orders and phase transitions among them. Transition metal dichalcogenide compounds serve as a bedrock for exploration of such hybrid systems. They host a variety of exotic electronic phases and their Van der Waals nature enables to admix them, either by exfoliation and stacking or by stoichiometric growth, and thereby induce novel correlated complexes. Here we investigate the compound 4Hb-TaS$_2$ that interleaves the Mott-insulating state of 1T-TaS$_2$ and the putative spin liquid it hosts together with the metallic state of 2H-TaS$_2$ and the low temperature superconducting phase it harbors. We reveal a thermodynamic phase diagram that hosts a first order quantum phase transition between a correlated Kondo cluster state  and a flat band state in which the Kondo cluster becomes depleted. We demonstrate that this intrinsic transition can be induced by an electric field and temperature as well as by manipulation of the interlayer coupling with the probe tip, hence allowing to reversibly toggle between the Kondo cluster and the flat band states. The phase transition is manifested by a discontinuous change of the complete electronic spectrum accompanied by hysteresis and low frequency noise. We find that the shape of the transition line in the phase diagram is determined by the local compressibility and the entropy of the two electronic states.  Our findings set such heterogeneous structures as an exciting platform for systematic investigation and manipulation of Mott-metal transitions and strongly correlated phases and quantum phase transitions therein.
}

Summary sentence: Spectroscopic imaging of a first order quantum phase transition between a Kondo cluster and a depleted Mott state in 4Hb-TaS2 

\section*{Main}

The spatial localization of electrons amplifies their electronic interactions and hence gives rise to strongly correlated phases with exotic properties \cite{Balents2020,Kennes2021}. One of the fundamental correlated states is the Mott-insulator where onsite Coulomb repulsion prevents double electron occupancy and gives rise to a charge arrest with no electron hopping \cite{Imada1998}. The quenched kinetic energy of the Mott-insulator is a bedrock for exotic electronic phases, magnetic orders and spin liquids \cite{Savary2017,Broholm2020}. Doping the Mott-insulator away from half filling induces additional correlated states with a rather universal, though still debated, phase diagram \cite{Lee2006}. Upon doping, magnetically ordered phases transition into a pseudo-gapped state before high temperature superconductivity settles and vanishes at yet higher doping levels. An interesting route to enrich this phase diagram and access novel electronic orders is by coupling the Mott insulator to distinct electronic states. The coupling to a band insulator has been shown to induce an interface metallic layer that in some cases even exhibits superconductivity as in oxide interfaces \cite{Reyren2007,Liu2021KTO}. Even the coupling of two distinct Mott-insulators was shown to induce interface superconductivity \cite{Misawa2016}, as well as, high temperature interface superconductivity, which can be manipulated by
modifying the oxidization conditions and doping \cite{Ju2021}. Here we investigate the response of coupling a  Mott-insulator to a metallic layer. We show that this hybridization, where both the charge and the spin degrees of freedoms interact, gives rise to an electronic phase diagram that hosts a novel first order quantum phase transition distinguishing between two electronic states \cite{Helmes2008, Rizzo2020}.  

To that end we study the transition metal dichalcogenide (TMD) 4Hb-TaS$_2$, which consists of alternating layers of 1T-TaS$_2$ and 1H-TaS$_2$. A monolayer of 1T-TaS$_2$ is a prototypical Mott insulator and a candidate quantum spin liquid \cite{Sipos2008,Law2017,Chen2020}. It hosts a triangular charge density wave (CDW) order under which clusters of 13 Ta atoms reconstruct in a star of David arrangement. This leads to the hybridization of 12 electronic states leaving a single unpaired electronic state per CDW site. Those unpaired states are believed to exhibit large onsite interaction, $U_{1T}$, compared to their hopping amplitudes and hence form a Mott insulator. Accordingly, the typical density of states (DOS) spectrum in 1T-TaS$_2$ shows an energy gap of about $U_{1T}\approx$ 400 meV appearing symmetrically across the Fermi energy. The energy gap is bounded by two DOS peaks, signifying the lower and upper Hubbard bands. This gives rise to charge arrest and arguably to a spin liquid phase \cite{Law2017,Chen2020}, as no magnetic order has been identified in pure 1T samples down to the lowest measured temperatures \cite{ribak2017}. There are conflicting indications, however, that in its bulk form hybridization of stacked CDW sites on neighboring 1T-TaS$_2$ layers results in the formation of a band insulator rather than a correlated one \cite{Ritschel2015,Ritschel2018,Lee2019,Butler2020}. 1H-TaS$_2$ is a metal and at low temperatures becomes a strongly spin-orbit coupled Ising superconductor \cite{DelaBarrera2018}. The combination of 1T and 1H polytypes thus provides a rich playground to explore strongly correlated effects \cite{Helmes2008,Al-Hassanieh2015}. This opportunity is readily available in single crystals of 4Hb-TaS$_2$, whose crystal structure is shown in Fig.\ref{fig1}a. 
  
In the hybrid compound both the superconducting and the Mott states are profoundly altered. The 4Hb-TaS$_2$ superconductivity was shown to onset concurrently with an increase in Muon spin relaxation \cite{Ribak2019}, suggesting  a chiral superconducting phase. More recent STM studies, visualized the existence of topological nodal superconductivity on the 1H surface termination of single crystals  \cite{Nayak2021} possibly akin to similar observations in MBE grown 1H monolayers \cite{Vano2021b}. The fate of the Mott insulating state on the 1T layers has thus far remained ambiguous. Spectroscopic STM measurements on 4Hb-TaS$_2$ single crystals have found that inter-layer hybridization results in shifting the Hubbard bands to above the Fermi energy \cite{Wen2021}. In contrast, the dI/dV spectrum of MBE grown 1T/1H bilayers exhibits clear signatures of Kondo behaviour  \cite{Vano2021,Ruan2020} due to the screening of the localized spins on the CDW sites by the metallic 1H layer underneath \cite{Helmes2008}, possibly constituting a Kondo lattice. A uniform Kondo lattice was also induced and manipulated by intercalating single crystals with Pb atoms \cite{shiwei22}.

Yet here, we report that both ground states - the Kondo states and the flat band - not only coexist in the 4Hb-TaS$_2$ sample but can even be reversibly switched on demand through a discontinuous transition. On the majority of the CDW sites we observe the upward shift of the Hubbard bands in energy. We show in calculation that this is accompanied by a charge transfer from the 1T to the 1H layers, leading to the  formation of a depleted doubly degenerate flat band (FB) above the Fermi energy. Yet, on about 12\% of the CDW sites, charge transfer is prevented and the spectrum exhibits a sharp  Kondo-like resonance. This results in a residual cluster of Kondo sites embedded among the surrounding flat band state. We show that this Kondo cluster state undergoes a reversible first-order quantum phase transition into the flat band state by variation of temperature and electric field as well as with the aid of the STM tip. We characterize the two ground states, the quantum phase transition that separates them and its evolution across the temperature, electric field and tip-sample interaction axes of the emergent electronic phase diagram. Below, we first characterize the two distinct ground states as they appear in the sample and then demonstrate the induced phase transition among them and investigate its thermodynamic quantum nature.

\section*{Results}

\subsection*{Depleted Mott versus Kondo states}

\begin{figure}[ht]
\centering
\includegraphics[width=0.98\linewidth]{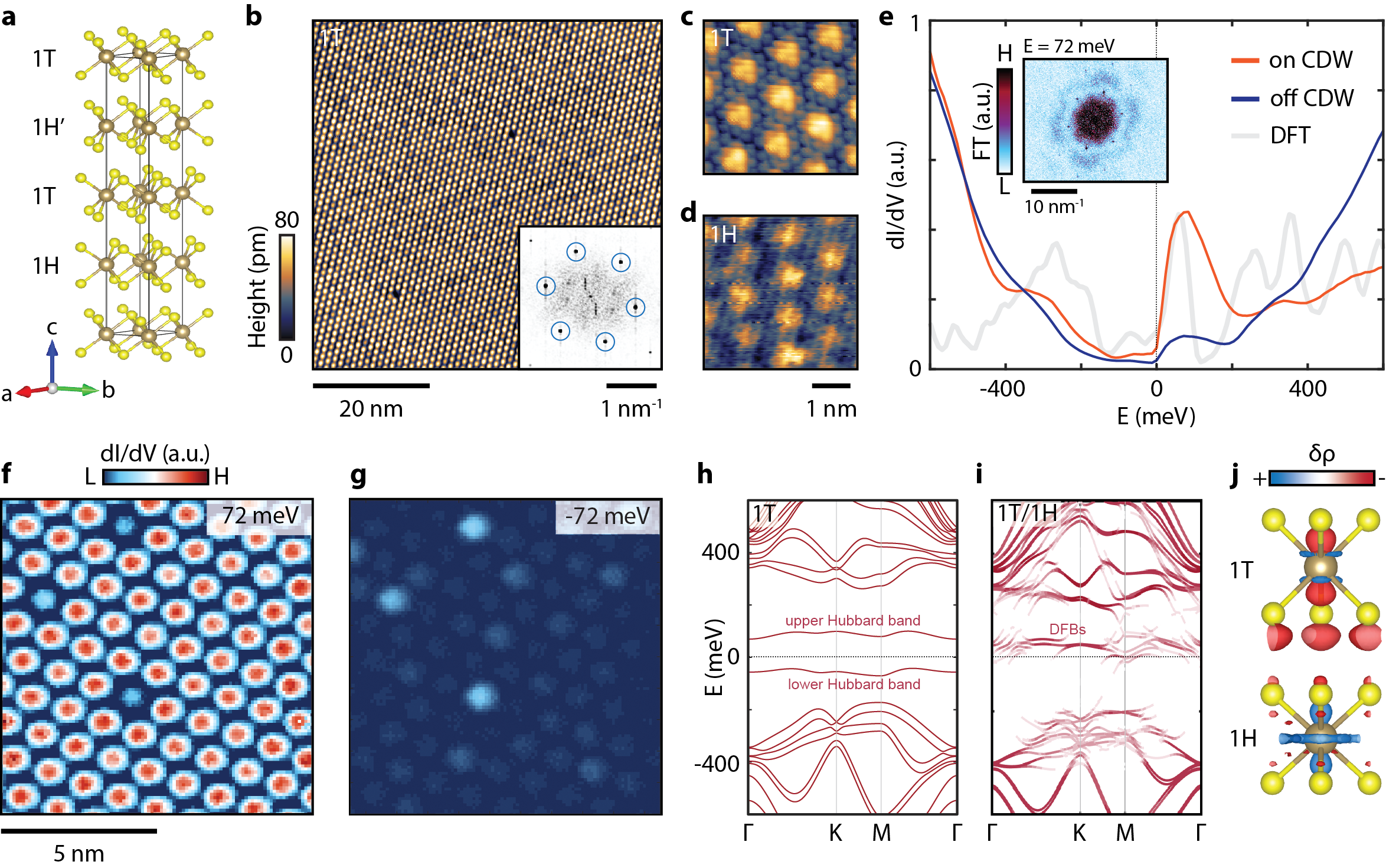}
\caption{\textbf{Depleted flat bands on the 1T-layer termination.} \textbf{a,} Unit cell of 4Hb-TaS$_2$ with alternatively stacked 1T and 1H layer coupled by van der Waals interaction. \textbf{b,} Large scale topographic image of the 1T layer showing commensurate CDW order. 10\% of the CDW sites are dim while the majority are bright. Inset: Fourier transform of the topography shows the $\sqrt{13}\times\sqrt{13}$ CDW peaks (blue circle). \textbf{c,} Atomically resolved topography of the 1T layer \textbf{d,} and of the 1H layer, both showing the $\sqrt{13}\times\sqrt{13}$ CDW pattern. \textbf{e,} Spatially averaged dI/dV profile measured on (orange line) and in between (blue line) 1T CDW sites. A robust peak is observed at about 75 meV on CDW sites. The spectrum agrees well with ab initio calculation of the integrated DOS (gray line). \textbf{f,g,} Spatially resolved dI/dV map of the 1T layer at $\pm$72 meV. \textbf{h,i,} ab initio calculations of a monolayer 1T showing the Mot-Hubbard state and 1H-1T bilayer showing the depleted flat bands \textbf{j} The depletion of the flat bands is facilitiated by charge transfer from the 1T to the 1H layer. Scanning parameters for \textbf{b-g}: $V_{set}$ = 600 mV, $I_{set}$ = 400 pA, $V_{ac}$ = 10 mV, f = 433 Hz. a.u., arbitrary unit.}
\label{fig1}
\end{figure}

Large single crystals of 4Hb-TaS$_2$ were grown with 1\% selenium added to the mixture (see Methods). They were mechanically cleaved at room temperature in ultra-high vacuum conditions and cooled down immediately in a commercial Unisoku STM. The cleaving process randomly exposes distinct 1T and 1H surface terminations. A typical large-scale topography of the 1T layer termination is displayed in Fig. \ref{fig1}b. It shows a commensurate $\sqrt{13}a_0\times\sqrt{13}a_0$ CDW pattern \cite{Wu1989,Coleman1988} at $T\sim4.2$ K. The CDW pattern forms a triangular super-lattice as captured by the Fourier transformed topography (inset). The star of David arrangement, composed of 13 atoms, is clearly seen in the atomically resolved topography in Fig.\ref{fig1}c (more details in SM section \ref{smwf}). Intriguingly, in the 4Hb-TaS$_2$ polytype the CDW pattern characteristic to the 1T layer is embedded also onto the adjacent 1H-TaS$_2$ layer (Fig.\ref{fig1}d), indicating the mutual interaction between the two layers.

The differential conductance (dI/dV) spectrum of topmost 1T-layers of 4Hb-TaS$_2$ samples look markedly different from the electron-hole symmetric Mott gap seen in pure 1T-TaS$_2$ \cite{Ma2016,Cho2016,Qiao2017,Butler2020,Butler2021}. Instead, the common on-CDW dI/dV spectrum (Fig.\ref{fig1}e, orange) displays an electron-hole asymmetric single pronounced peak that onsets sharply above the Fermi energy and reaches a maximum at about 75 meV. The DOS averaged over the network region, in between CDW sites (blue line), is highly suppressed signifying that the charge distribution remains fairly localized on the CDW sites. The spatial distribution of the DOS and its electron-hole asymmetry  above and below the Fermi energy, are mapped in Fig.\ref{fig1}f and g, respectively (the three distinct CDW sites that appear dim at positive bias and bright at negative bias will be discussed below). 

We performed ab initio calculations to resolve the exact origin of the spectrum we observe on the CDW sites. We model a 4Hb-TaS$_2$ bilayer and compare it with the band structure of a 1T-TaS$_2$ monolayer. In the parent 1T-layer, shown in Fig.\ref{fig1}h, one finds the known half-filled Mott spectrum hosting two flat Hubbard bands residing symmetrically above and below the Fermi energy (more details in section \ref{fig:DFT_BS}). Upon introduction of an adjacent 1H-layer, the lower Hubbard band gets depleted and shifts in energy to above the Fermi level, giving rise to a pair of fully-depleted flat bands that reside above the Fermi energy. This is displayed in Fig.\ref{fig1}i that shows the weight of the 1T/1H bilayer DOS projected on the 1T layer (the complementary projection to the 1H layer is in Fig.\ref{fig:DFT_BS}c). The corresponding integrated DOS shows a sharp peak above the Fermi energy that agrees with the peaked spectrum we measure on the 1T layer (Fig.\ref{fig1}e, gray and orange lines, respectively).
  
This charge depletion from the lower Hubbard band associated with the 1T layer is achieved by charge transfer from the 1T layer to the metallic 1H-layer \cite{Wang2018}, as shown in Fig.\ref{fig1}j. This charge transfer to the 1H layer is evidenced by the 1T characteristic charge modulation that we commonly observe on the 1H layer (Fig.\ref{fig1}d). Intriguingly, the quasi-particle interference (QPI) pattern within the energy window of the flat bands (Fig.\ref{fig1}e, inset, and in section \ref{fig:QPI}) is symmetric about the CDW Bragg peaks, rather than the atomic ones, signifying the folding of the electronic wavefunction by the CDW superlattice. The measured signature of flat bands agree with previous measurements on single crystal 4Hb-TaS$_2$ \cite{Wen2021}, but is markedly different from the Kondo spectrum found in artificial 1T/1H bilayers grown via MBE \cite{Vano2021,Ruan2020} that we discuss next.

\begin{figure}[ht]
\centering
\includegraphics[scale=0.9]{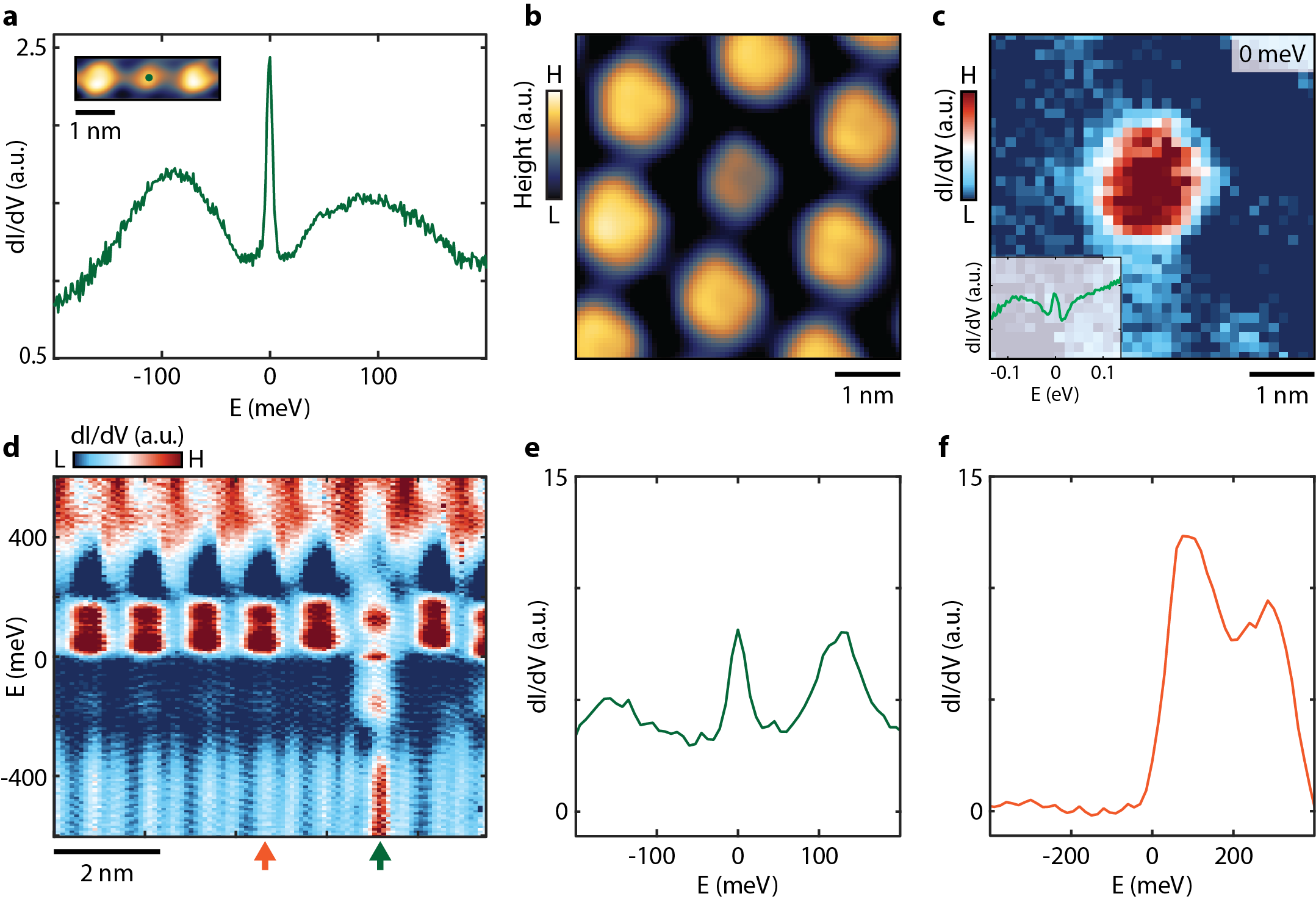}
\caption{\textbf{Critical Mott-like state on dim CDW sites.} \textbf{a,} A representative spectrum measured on a dim CDW site (topography in inset) exhibiting a sharp zero bias peak amidst two satellite peaks. Scanning parameters: $V_{set}$ = 200 mV, $I_{set}$ = 100 pA, $V_{ac}$ = 3 mV, f = 773 Hz. a.u., arbitrary unit. \textbf{b,} Topographic image of a dim CDW site amidst bright ones and \textbf{c,} corresponding dI/dV mapping at zero bias. \textbf{d,} dI/dV map measured on the 1T layer along a CDW axis. \textbf{e,f} dI/dV spectra taken over a dim and a bright spot marked by green and orange arrows in d, respectively. \textbf{g,} dI/dV map measured on the 1T layer along a CDW axis. \textbf{h,i} dI/dV spectra taken over dim and bright spots marked by green and orange arrows in g, respectively. Scanning parameters: $V_{set}$ = 600 mV, $I_{set}$ = 250 pA, $V_{ac}$ = 10 mV, f = 433 Hz.}
\label{fig2}
\end{figure}

We now focus on a subset of the CDW sites  that appear dim (shallower) in positive bias topography as shown in Fig.\ref{fig1}b. Those dim sites constitute about 12\% of all CDW sites (see SM \ref{fig:topo_dim}). Their shallow appearance is not topographic but rather rooted in their electronic spectrum. While their exact chemical origin is still uncertain, it may be related to incorporation of Se that was added to the mixture. The local DOS spectrum characteristic of such dim CDW sites is shown in Fig.\ref{fig2}a (see also Fig.\ref{fig:ZBP_1}). It displays a sharp zero bias conductance peak (ZBCP) in agreement with recent visualization of a Kondo response in MBE grown 1T/1H bilayers \cite{Ruan2020,Vano2021}. There, however, it appeared uniform across all CDW sites, while our single crystals mostly host the flat band states and the Kondo spectrum appears only on a subset of sites forming a rather dense cluster.  
The width of the ZBCP, that reflects the width of the coupling to the metal, is on the order of a few meV (see Fig.\ref{fig:fano_fit}) and somewhat varies between dim sites. The satellite peaks correspond to charge excitations
whose energy spacing, $U \approx 200$ meV, captures the onsite repulsion and their width, $W\approx 200$ meV, represents the coupling strength to nearby sites or to the metal. The ratio of the two scales, $U/W\approx$1, puts those Kondo state close to the intriguing critical regime. We find no response of the Kondo spectrum to application of the maximal magnetic field of 9 T that we can reach (see Fig.\ref{fig:ZBP_B}). While this is quite atypical to isolated Kondo impurities \cite{otte2008}, such robustness is rather common in Kondo lattices. In particular, it is consistent with observations made in the MBE grown 4Hb-TaS$_2$ bilayers that host a regular Kondo lattice and showed splitting of the Kondo ZBCP only above a field of 10 T \cite{Vano2021}. This suggests that the Kondo cluster that forms in our samples already exhibits Kondo lattice-like correlations. 

Spatial spectroscopic mapping over the topography shown in Fig.\ref{fig2}b of a similar dim CDW site positioned at its center, given in Fig.\ref{fig2}c, shows that the ZBCP (full spectrum shown in inset) is distributed across the dim CDW site rather than localized over a single atomic defect (see also section \ref{fig:ZBP_MAPS}). Spatial mapping of the spectrum along a row of CDW sites, shown in Fig.\ref{fig2}d (more in Fig.\ref{fig:ZBP_LC_2}), shows the stark distinction between the symmetric Kondo spectrum with satellite peaks on a single dim site to the asymmetric depleted flat band (DFB) state we find on all neighboring sites (spectra shown explicitly in Fig.\ref{fig2}e and f, respectively). Intriguingly, CDW sites adjacent to the dim one seem to be somewhat affected as their local spectrum slightly shifts in energy (see also \ref{fig:ZBP_LC_1}) suggesting that the local dim site potential has a long range effect.

\subsection*{First order quantum phase transition in the voltage-temperature phase diagram}

\begin{figure}[ht]
\centering
\includegraphics[width=\linewidth]{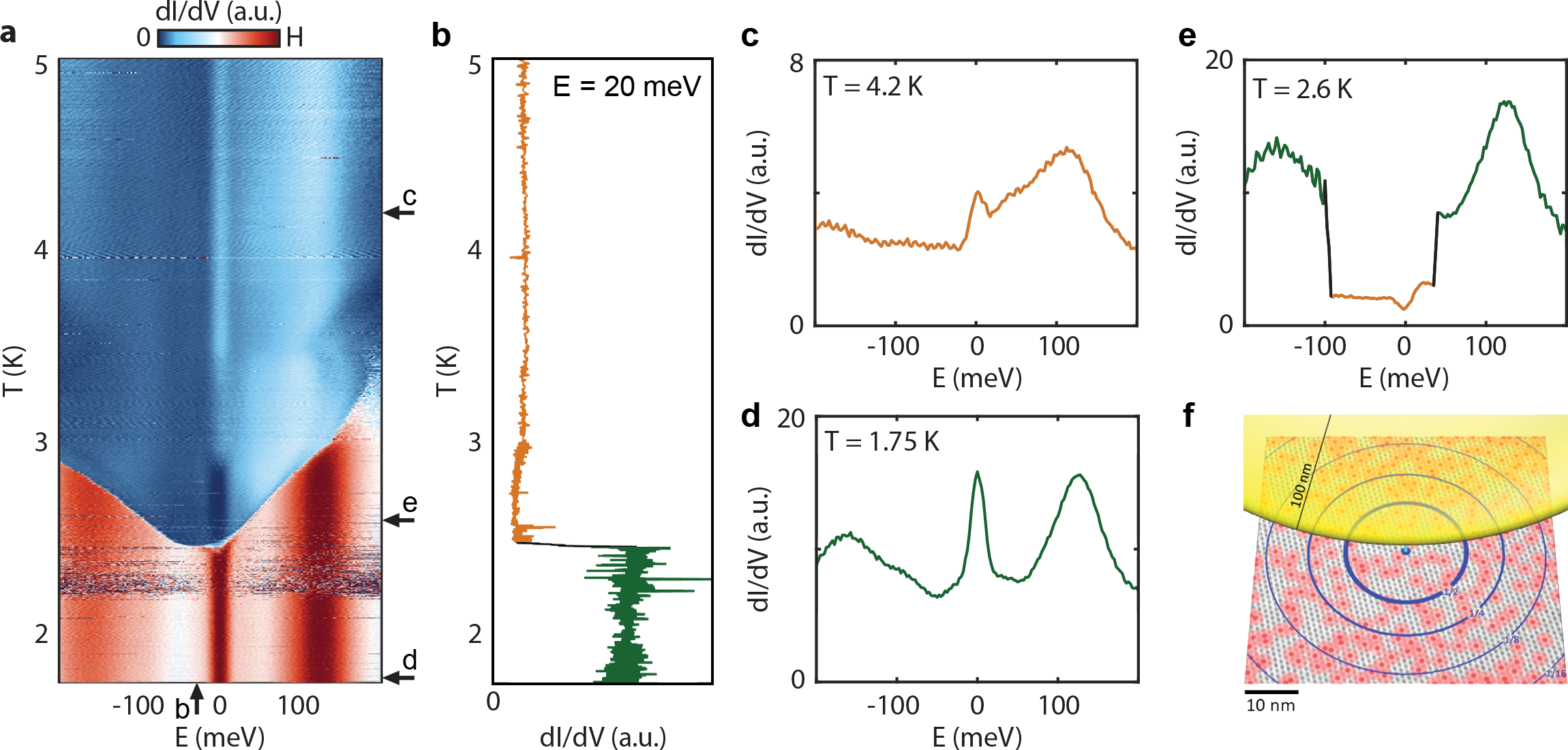}
\caption{\textbf{Temperature-Voltage induced phase transition} \textbf{a,} Temperature-bias voltage evolution of the dI/dV spectrum showing temperature-voltage sharp transition line.  \textbf{b,} Temperature cut at a fixed bias of E=20 meV showing the sharp transition. \textbf{c,} High temperature dI/dV spectrum showing asymmetric DFBs. \textbf{d,} Low temperature dI/dV spectrum showing symmetric Kondo spectrum.  \textbf{e,} At intermediate temperatures sharp discontinuity in dI/dV occurs both at positive and negative bias signifying the phase transition.  \textbf{f,} Red dots mark positions of dim defect sites over the same topography as in Fig.\ref{fig1}b and light red circles mark their nearest neighbors. The blue circles mark radii at which the electric field induced by the STM tip with 100 nm apex curvature positioned 1nm above the sample (yellow) is attenuated by consecutive factors of 2 from its maximal central value.}
\label{fig4}
\end{figure}

So far all the spectroscopic imaging was conducted at a fixed temperature. Next, we follow the evolution of the dI/dV spectrum on a dim CDW site with varying temperature and electric field and show that it undergoes an abrupt transition from a Kondo state into a flat band state. In Fig.\ref{fig4}a we show the dI/dV spectrum measured while gradually raising the temperature. Remarkably, the electron-hole symmetric Kondo spectrum we measure at low temperatures switches to the electron-hole asymmetric flat band spectrum as the temperature is increased. A cut across temperature at a certain applied bias (marked by arrow) is shown in Fig.\ref{fig4}b demonstrating the discontinuous character of the transition. At high temperatures, above 3 K, we find the electron-hole asymmetric spectrum of the flat band as shown in Fig.\ref{fig4}c. At low temperatures, below 2.5 K, the electron-hole symmetric Kondo spectrum takes over, as shown in Fig.\ref{fig4}d. The abrupt jump in the local DOS traced across temperature and electric field thus distinguishes the different electronic spectra of the different electronic ground states. It thus signifies a quantum phase transition line between the Kondo state, observed on dim sites and the flat band ground state associated with the bright sites. 

As the mapping in Fig.\ref{fig4}a shows the transition is not driven by temperature alone but by its combination with electric field. The latter is applied by the STM tip as the bias voltage is scanned. Indeed, voltage bias scans taken at a fixed temperature within the intermediate range of 2.5-3 K show two sharp discontinuous steps in dI/dV, occurring at positive and negative bias as demonstrated in Fig.\ref{fig4}e. These sharp steps in dI/dV that gradually shift towards zero bias with decreasing temperatures both signify the transition from a Kondo state at high negative and positive bias to the flat band state at lower bias. The observation of a sharp transition between distinct dI/dV spectra whose associated temperature varies with applied electric field is remarkable. We could only resolve the evolution of the energy scale at which it occurs from 0 to 200 meV over a temperature window of 0.5 K across which the transition remains perfectly abrupt within the framework of a phase transition. Not all dim CDW sites exhibit such temperature-field driven transition among the parameter spaces we covered. Nevertheless, the occurrence of the phase transition on a subset of the sites allows us to uniquely investigate the nature of the quantum phase transition and its evolution across the mapped phase diagram. 

We stress, that unlike the tip to sample tunneling that decays exponentially over a length scale of picometers the electric field the tip imparts is quasi-local as it decays as a power-law. Accordingly, the extended geometry of the STM tip, with a typical radius of curvature of 100 nm (sketched to scale in Fig.\ref{fig4}f) should be considered \cite{pasupathy2021}. The blue circles in Fig.\ref{fig4}f mark radii at which the electric field between a half-sphere and a plane decays by consecutive multiples of 2 from its peak intensity at the center. It thus becomes evident that while we probe with tunneling the site immediately under the tip, its electrostatic coupling to the sample extends over hundreds of CDW sites.

We use the transition line to further characterize the thermodynamic nature of the first-order quantum phase transition and the competing energy scales involved in it. The transition line appears fairly symmetric across positive and negative bias in the mapped bias-temperature phase diagram. As discussed above, ab initio calculation suggests that the transition involves charge transfer between the localized 1T CDW sites and the 1H metal. This varies the strength of the electric dipole, $\mathbf{P}$, between the charge and the STM tip. The coupling energy of the electric field imparted by the STM tip, $\mathbf{E}$, to that dipole, $E_1= \mathbf{P}\cdot \mathbf{E}$, contributes to the free energy of the system and could have driven charge transfer. However, the linear coupling of the electric field to that electric dipole cannot account for the quadratic shape of the transition line seen in Fig.\ref{fig4}a. 
This suggests that the phase transition is driven by the charging energy $E_2 = -\frac{1}{2} C^i U^2$ which is quadratic in the bias voltage, $V$. Since the geometric capacitance is essentially fixed by the shape of the tip and the underlying metallic layer, the transition is rather driven by the contribution to the energy that arises from the difference in the local copmpressibility between the electronic phases on either side of the transition line. These energetic considerations compete with the entropic ones, $F^i=E_1^i+E_2^i-TS^i$, where $i=1,2$ corresponds to the phases on either side of the transition. The thermodynamic first order transition occurs at a voltage dependent temperature, $T_{QPT}(V)$, obtained from $F^1(V,T_{QPT})=F^2(V,T_{QPT})$.  Therefore, we conclude that the Kondo state that appears at lower temperatures is favored due to its higher compressibility relative  to the flat band state that is favored by entropy at higher temperatures. 

The higher compressibility of the Kondo spectrum naturally arises from the metallic ZBCP compared to the insulating depleted flat bands. The origin of excess entropy of the flat bands at high-temperature is more elusive. Most electronic phase transitions show exactly the opposite behavior: typically low-temperature phases  have a reduced instead of an enhanced DOS at the Fermi energy. Thermodynamically, the high-temperature phase has to have a higher entropy compared to the low-temperature phase, while our spectroscopy of the depleted band reveals that the excess entropy cannot arise from electronic excitations in the top 1T layer. Most likely, the excess entropy therefore has its origin in fluctuating spin or charge degrees of freedom which arise when charge is transferred from the 1T to the 1H layer and the detailed nature of the hybridized state that forms as a result of it. It could also arise from the exact character of the microscopic origin of the dim sites, that we could not resolve with certainty. 

Though subdominant relative to the compresibility, the charge transfer associated with the phase transition still induces an electric polarization which is reflected in the slight offset of the minimum of the quadratic transition line from zero bias. However, the small value we obtain for the offset of about 25 meV signifies that either the polarization is far smaller than what geometrical considerations suggest or that it is balanced off by a large entropy gain of $10^2-10^3 k_B$ (see section \ref{smthermo}). The origin for the screening of the dipole or excess entropy is yet unresolved though in part it could be related to the relative orientation of the dipole and local electric field that enter the free energy as a vector product. Again, they may also arise from the detailed nature of the defect sites or the hybridized state that they form with the 1H metal that require further investigation.

\subsection*{Tip-induced quantum phase transition}

We next show that the quantum phase transition can be induced also by modifying the 1T-1H interlayer coupling via tuning of the distance between the sample and the STM tip and use it to further characterize its nature. In the temperature-bias scans we have discussed above the tip was used as a local electrode \cite{Dombrowski1999,Zhao2015,Collins2018,pasupathy2021}. Other than that, tip  invasiveness did not play a role in inducing the transition as the tip position was kept mostly fixed (one exception happens when the phase transition line crosses the parking bias as discussed in section \ref{fig:vtraw}). More generally though, the STM tip may have different functionalities besides enabling spectroscopic imaging. For instance, it may provides an extra screening channel thus modifying local Coulomb potentials and it may interact with the sample via inter-atomic forces that may alter the inter-layer coupling of layered materials \cite{Yankowitz2016,Georgi2017}. We show that on certain dim sites the tip's invasiveness too induces the reversible quantum phase transition  between the Kondo and flat band states. 
the STM tip height above the sample is set not only by the surface topography but also by the integrated DOS from the Fermi energy up to the set bias voltage. 

\begin{figure}[ht]
\centering
\includegraphics[width=\linewidth]{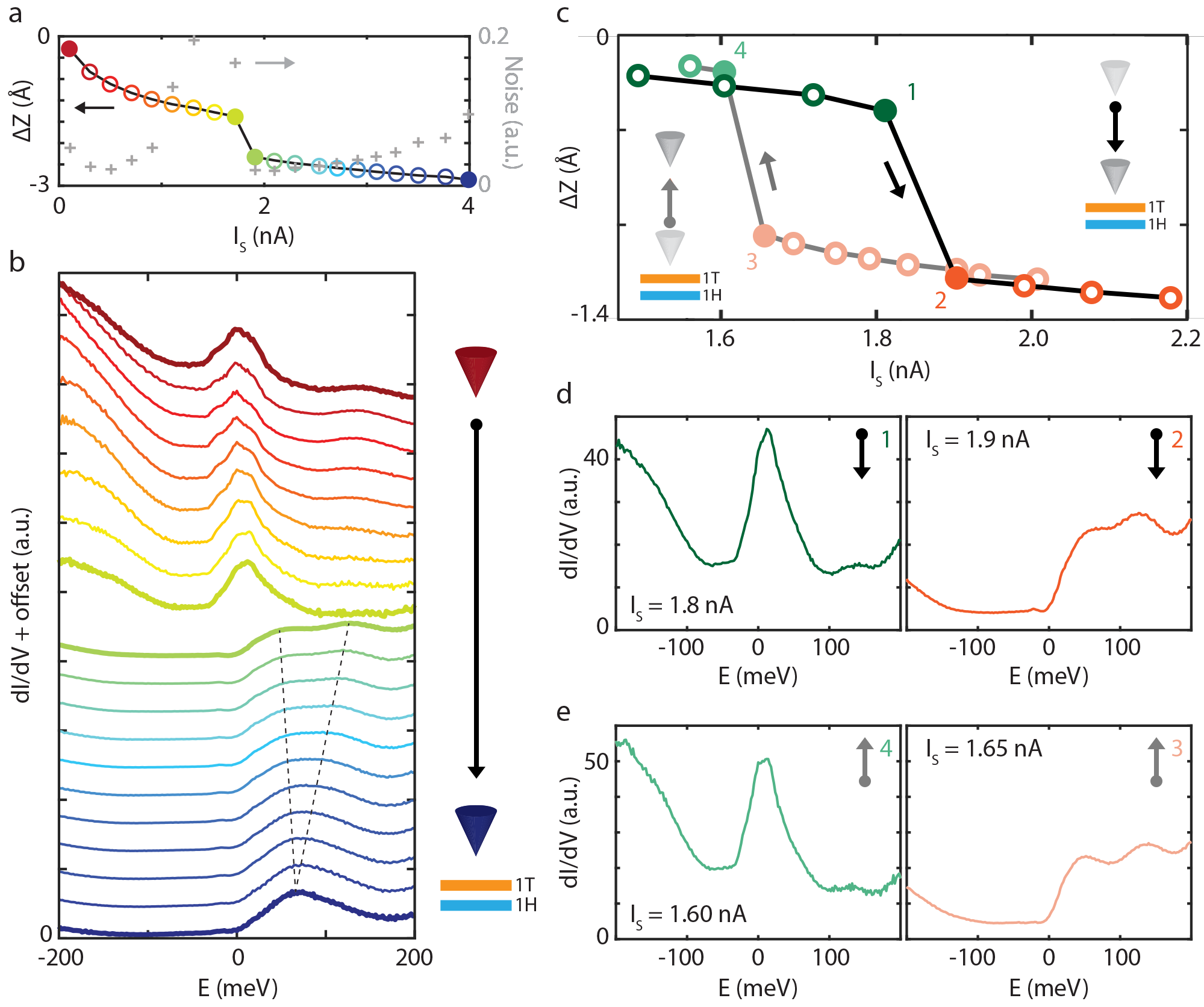}
\caption{\textbf{Tip induced phase transition.} \textbf{a,} Tip height displacement versus tunneling current under 200 meV bias voltage \textbf{b,} Consecutive dI/dV spectra measure at corresponding height in a showing an abrupt jump from Mott-like to flat band spectra. \textbf{c} High resolution tip approaching (green) and retracting (orange) sweeps showing hysteresis in transition point. \textbf{d,e,} individual spectra taken on approaching (green) and retracting (orange) sweeps just before and right after transition point (on corresponding points numbered in c). }
\label{fig3}
\end{figure}

We position the tip at a certain height (on the order of a nanometer) above a dim CDW site by stabilizing a tunneling current, $I_s=$0.2 nA, under a chosen bias voltage, $V_b=$200 meV (corresponding to the solid red circle in Fig.\ref{fig3}a). The spectrum measured at this initial height shows a Kondo type spectrum hosting a ZBCP (thick red line in Fig.\ref{fig3}b) already at 4.2 K - indeed, we find variability at the local conditions at which the transition occours that we attribute to inhomogeneity. We then bring the tip closer to the sample by increasing the set current in steps of $\Delta I_s$ (Fig.\ref{fig3}a) and monitor the evolving spectrum as the tip approaches the dim CDW site (Fig.\ref{fig3}b). Over the initial approach of about 1.5 Angstroms we find no change in the Kondo spectrum. Then, over a single additional step of $\Delta I_s$ the spectrum switches abruptly from the symmetric Kondo-type to the asymmetric flat band type. The changing spectrum results in a slightly larger tip height repositioning step over the otherwise exponential background (Fig.\ref{fig3}a). We use it as a convenient proxy for identifying the point of transition as it reflects a corresponding change in the spectrum (Fig.\ref{fig3}b). 

Just past the transition the doubly degenerate flat bands are slightly split in energy, as can be deduced from the double peak structure that characterizes the spectrum. As we continue to approach with the tip further towards the sample the two peaks gradually merge (tracked by dotted lines) into a single peak, resulting in an identical spectrum to the asymmetric degenerate flat band spectrum we commonly find on bright CDW sites (as in Fig.\ref{fig1}e). This transition in response to the tip position is  similar to the one we have demonstrated above in response to temperature and electric field induced by the tip. Bringing the tip closer amplifies the electric field it induced at a given bias voltage. However, in response to increasing electric field we have induced the opposite transition from a flat band to a Kondo spectrum (see Fig.\ref{fig4}e). We thus conclude that the leading effect of bringing the tip closer to the sample is the increasing atomic forces between the two, which in turn affects the coupling between the vand der Waals 1T and 1H layers. Tip invasiveness thus adds a third axis to the temperature-bias phase diagram we have presented above (Fig.\ref{fig4}a).

As the transition point is approached we also find a concurrent rise in the noise level of the tunneling dI/dV spectrum. It is seen directly in the individual spectra in Fig.\ref{fig3}b and captured by the standard deviation in dI/dV about its mean value shown by plus symbols in Fig.\ref{fig3}a. The noise increases sharply on the Kondo branch towards the transition point (regardless of direction of height scan, see section \ref{fig:Hysteresis}) and reduces back upon crossing the transition point. The noise spectrum captured here is of low-frequency with a cutoff of about 1 KHz. Intriguingly, the spectra acquired within the Kondo state appear noisier also in the temperature-bias mappings (see for instance Fig.\ref{fig4}b). Seldomly we have also captured telegraphic noise over dim CDW sites with comparable typical switching time-scale of tens of milliseconds up to seconds (Fig.\ref{fig:noise}) that was absent on bright sites. Such low frequencies are far from being characteristic of single electron processes \cite{Aishwarya2019} which are typically on the order of terahertz. This suggests that spatial correlations are present on the Kondo side of the transition that thus involves a larger cluster rather than a single CDW site.

We further examine the sharp transition and its reversibility by taking finer steps about it through the approaching and retracting branches, shown in Fig.\ref{fig3}c by dark and light lines, respectively (the corresponding dI/dV spectra are in Fig.\ref{fig:Hysteresis}). Even at such high resolution we still we find that the transition between the Kondo and the flat band spectra occurs abruptly. Remarkably though, the transition height shows a clear hysteresis loop between approaching and retracting branches (Fig.\ref{fig3}d and e). These observations too indicate that the Kondo and flat band spectra do not evolve continuously from one to another, but are rather distinguished by a first-order quantum phase transition. It is triggered by the charge transfer back and forth from the 1T-layer to the metallic reservoir of the 1H-layer that can be induced reproducibly by the STM tip (see Fig.\ref{fig:ZBP_IZ}) as well as by temperature and bias. Not all CDW sites that host Kondo states allow such tip manipulation within the parameter space we have tested. Nevertheless, the sites on which such tunability is realized enable us to uniquely investigate the interplay between the two electronic states. The origin of the hysteresis may involve the tip control as the abrupt change in the spectrum is accompanied with an abrupt repositioning of the tip height. No similar response was detected when repeating such tip approach protocols on bright sites hosting flat band spectrum (see section \ref{fig:IZ}). 
                                
\section*{Discussion}

The appearance of a first order quantum phase transition with such temperature-bias relation who is accompanied by a hysteresis loop and associated low frequency noise does not adhere to the phenomenology of an isolated electronic defect state. The typical time scales of electronic states of matter and specifically switching times of isolated Kondo states are incompatible to those we encounter \cite{Delattre2009}. The lack of magnetic response of the ZBCP is also characteristic of correlated Kondo sites as in a Kondo lattice \cite{Vano2021} rather than isolated Kondo states \cite{otte2008}.  Indeed, 12\% concentration of dim defect sites translate to dim sites being next nearest neighbors on the average. The connected cluster is demonstrated on top of the topographic image in Fig.\ref{fig4}f, where all dim defect sites are marked with a red dot and the distance to their nearest neighbors by light red circle showing the close proximity. On the other hand, the electrostatic coupling of the STM tip to that Knodo cluster extends over hundreds of CDW sites \cite{pasupathy2021} as demonstrated in Fig.\ref{fig4}f by concentric circles. The imaged phase transition thus signifies the presence of correlations at least on a mesoscopic scale. 

The observations reported here provide a potential exciting pathway for investigating the gradual onset of correlated phases and phase transitions among them. However, to perform this systematically the origin of the Kondo sites has to be clarified. If we assume a dim CDW site is caused by a single atomic defect, their abundance per atom (12\% of all CDW sites with 13 atoms in each site) roughly agrees with the 1\% selenium to sulfur concentration used during growth for stabilizing the 4Hb polytype of TaS$_2$. Indeed, compared to 1H-TaS$_2$, the work function of 1H-TaSe$_2$ is slightly closer to that of 1T-TaS$_2$. However, quantitatively to prevent charge transfer to occur in ab initio calculation requires substituting all 13 sulfur atoms under a certain CDW site with tellurium rather than selenium atoms, neither of which are realistic. Similarity with recent imaging of Pb intercalated 1T-TaS$_2$ \cite{Shen2021} may suggest that selenium is intercalated between the 1T and 1H layers rather than substituted. On the other hand, the phenomenology that we report here seems to be dissimilar to a recent study of Pb intercalation that induced a uniform Kondo lattice that allowed some tunability but did not undergo a phase transition \cite{shiwei22}.  We have never observed a clear atomic defect nor an adatom on the 1T or 1H layers. If indeed the Kondo site concentration is set by Se concentration in the flux it would be exciting to gradually modify it in future investigations and explore the evolution of the electronic phase diagram and the quantum first-order phase transition it hosts.

Our spectroscopic findings also reflect on previous observations reported on 4Hb-TaS$_2$. The identification of a dense Kondo cluster, where the spin on the 1T layer is screened by the metallic 1H layers, could suggest alternative origin for the enhancement of Muon spin relaxation that was seen to onset once the material turns superconducting \cite{Ribak2019}. Within the superconducting state Kondo screening is suppressed, thus revealing a stronger magnetic environment for the Muons. More recently, an intriguing vortex memory effect from above the superconducting critical temperature was identified in single crystals of 4Hb-TaS$_2$ \cite{beena2022}. It suggests the existence of a time-reversal symmetry broken phase that lacks magnetization but still facilitates magnetic memory. It was attributed to a chiral spin liquid phase that has been considered to reside on the 1T-TaS$_2$ layers. However, our observations of charge and spin transfer from the majority of CDW sites on the 1T-TaS$_2$ layers to the 1H-TaS$_2$ metal (superconducitng below 2.7 K), alongside the spectroscopic collapse of the Mott spectrum, exclude the formation of a spin liquid phase there. Nevertheless, our identification of a Kondo cluster may provide alternative context for future investigations for the combined ground state, such as a Kondo glassy state, and its effect on the superconducting state at lower temperatures.

Intriguingly, the zero-bias transition temperature of the quantum phase transition we report here coincides with the superconducting transition temperature of bulk 4Hb-TaS$_2$. While this may be a mere coincidence, it can also hint that the low temperature Knodo cluster state is more susceptible to condensation and thus promotes a concurrent superconducting transition \cite{senthil2003}. With respect to previous spectroscopic reports of 4Hb-TaS$_2$, our observations here suggest that the mechanism that led MBE grown 1H/1T bilayers to show a uniform Kondo state \cite{Vano2021,Ruan2020} lies in absence of charge transfer. A plausible reason for that is lack of sufficient commensurate conditions that is required to induce it. In contrast, pure 4Hb-TaS$_2$ single crystals do not host defect states \cite{Wen2021} that have allowed us here to induce and manipulate the first-order quantum phase transition that separates the two states. 

\clearpage
\section*{Figures}

\clearpage

\clearpage

\clearpage

\clearpage
\section*{Methods}

\subsection*{DFT calculations}
We performed the density-functional theory (DFT) calculation follows in the framework of the generalized gradient approximation \cite{Perdew1996}) with the Vienna \textit{ab-intio} package \cite{Kresse1999}. We employed the PBE-D2 method to describe vdW interaction \cite{Grimme2006}. We included spin-orbit coupling (SOC) interaction in all calculations. The 4H$_b$ structure with $\sqrt{13}\times\sqrt{13}$ supercell was used for the CDW reconstructed structure. Full geometry optimization was performed until the Hellmann-Feynmann force acting on each atom became smaller than 0.01 eV/\AA. 

In the optimized structure, the 1T layer forms the star of David (SD) structure and the 1H layer remains almost the same (less than 1\% change). To evaluate interlayer interaction, we evaluated the charge transfer between the 1H and 1T layers. Because the 1T phase has about half eV lower work function than the 2H phase \cite{Shimada1994}, charge may transfer from the 1T layer to the 1H layer when forming an interface. As the result of charge transfer, We indeed observe the decrease (increase) of charge in the 1T (1H) side, resulting a charge dipole at the 1T-1H interface. 

\section*{Data availability}
The data that support the plots within this paper and other findings of this study are available from the corresponding authors upon reasonable request.

\printbibliography[title=References]

\section*{Acknowledgement}
We acknowledge fruitful discussions with Jonathan Ruhman, Erez Berg, Yuval Oreg and Ady Stern. HB and AR acknowledge funding from the German Research Foundation (DFG) through CRC 183 (project number 277101999, A01 and A02). BY acknowledges the financial support by the European Research Council (ERC Consolidator Grant, No. 815869) and the Israel Science Foundation (ISF No. 3520/20).

\section*{Author contribution}

\section*{Competing interests}
The authors declare no competing interests.

\section*{Supplementary information}
The supplementary information file contains sections 1-14 and figures S1-S23.



\include{SM/SM}

\end{document}

%% file: SM/SM.tex
\renewcommand{\figurename}{\textbf{Fig.}}
\renewcommand{\thefigure}{S\arabic{figure}}

\renewcommand{\thesection}{S\arabic{section}}
\setcounter{figure}{0}

\title{Supplementary Information for 'First Order Quantum Phase Transition in the Hybrid Metal-Mott Insulator Transition Metal Dichalcogenide 4Hb-TaS$_2$'}
\maketitle

\begin{refsection}

\section{Wave function distribution} \label{smwf}
\begin{figure}[ht]
\centering
\includegraphics[scale=1]{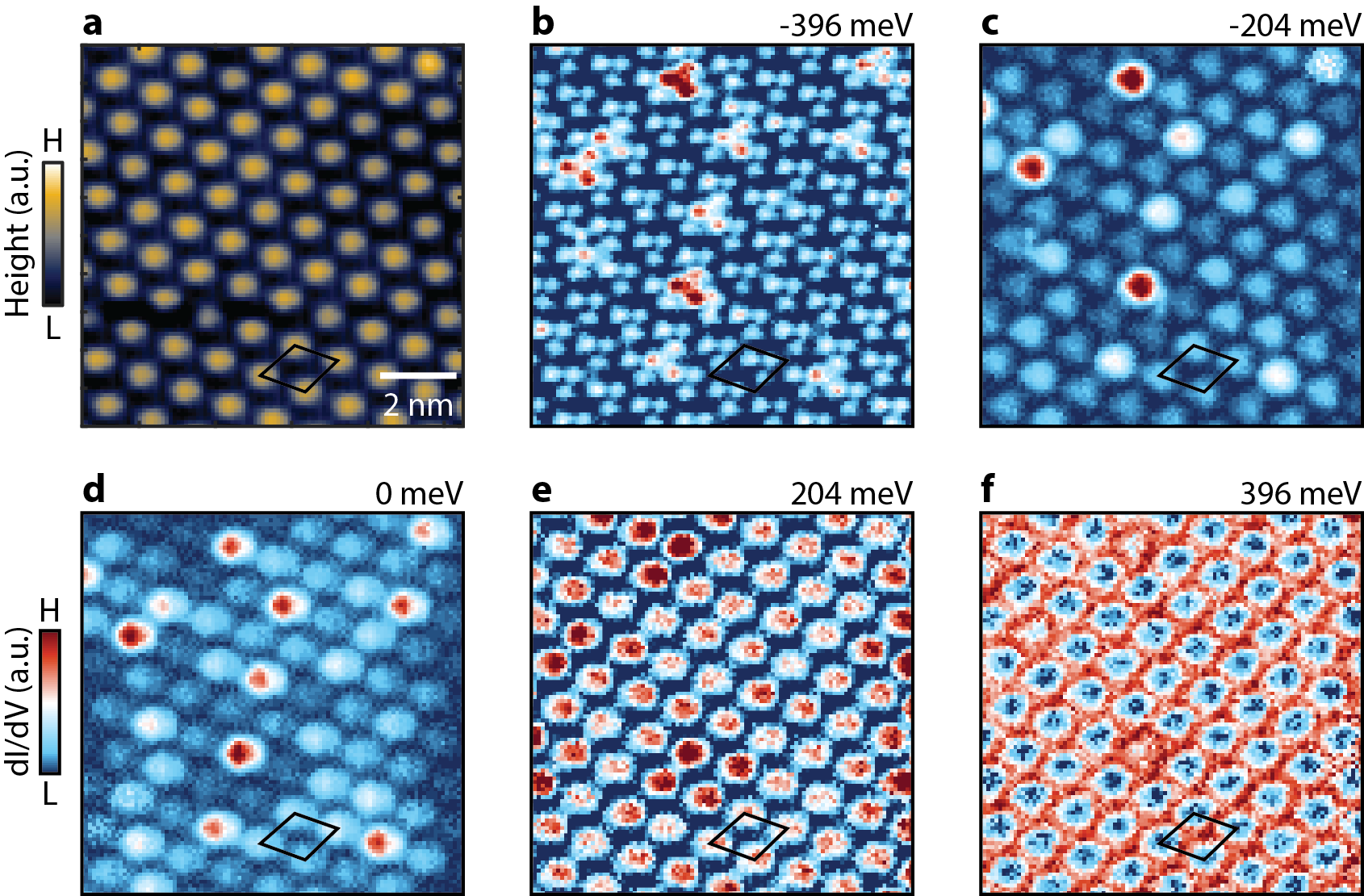}
\caption{\textbf{Wave-function mapping of the 1T layer.} \textbf{a,} Topography of the 1T layer measured along with the spectroscopic map. \textbf{b-f,} Spatially resolved spectroscopic map measured on \textbf{a} showing the evolving wave function distribution. Scanning parameters: $V_{set}$ = 600 mV, $I_{set}$ = 400 pA, $V_{ac}$ = 10 mV, f = 433 Hz. a.u., arbitrary unit.}
\label{fig:wfn_map_1}
\end{figure}

\begin{figure}[ht]
\centering
\includegraphics[scale=1]{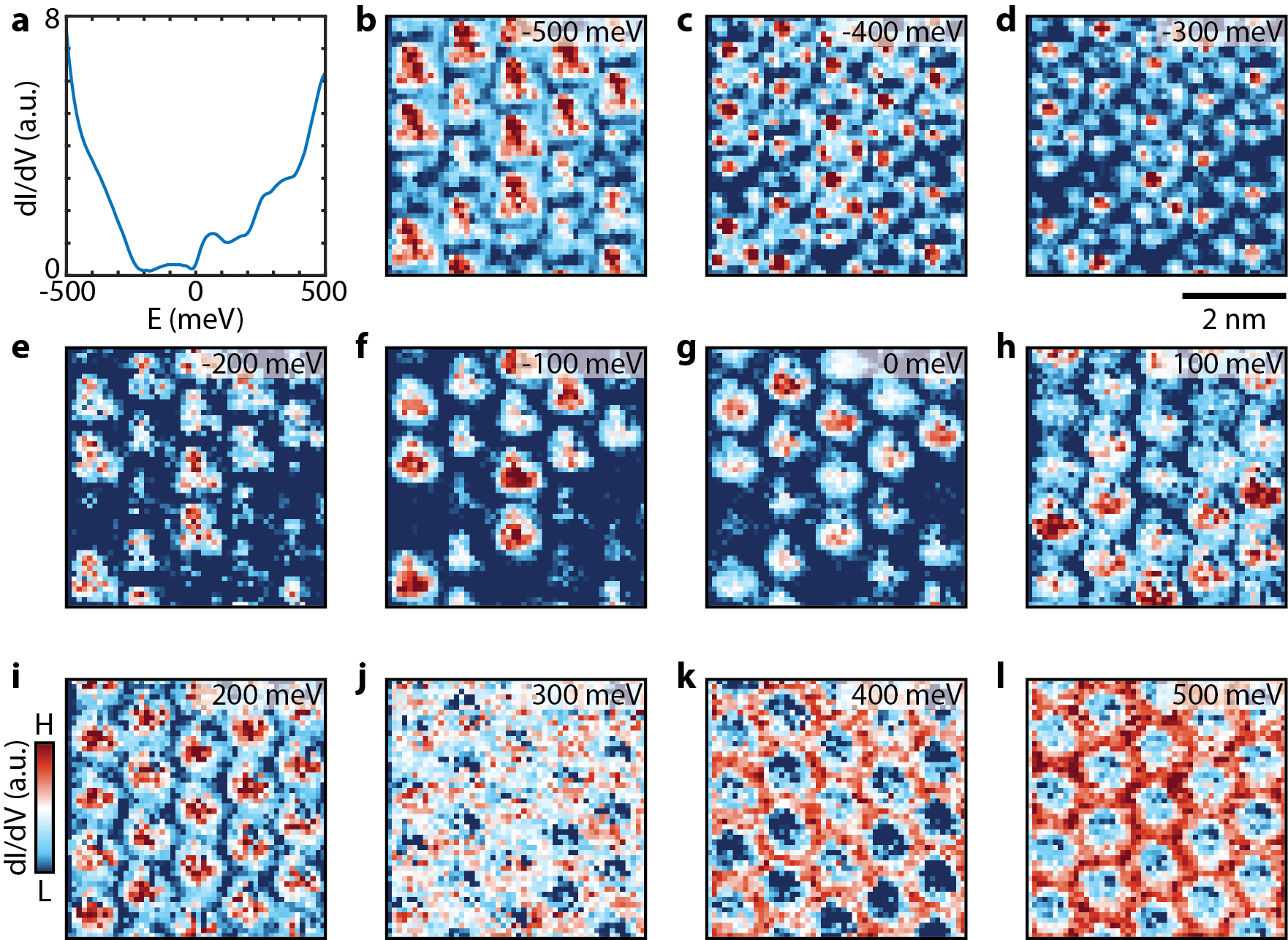}
\caption{\textbf{Additional wave-function mapping of the 1T layer.} \textbf{a,} Spatially averaged spectroscopy on the 1T layer. \textbf{b-l,} Spatially resolved spectroscopic map measured on the 1T layer showing the evolving wave function distribution similar to Fig. 1\textbf{d-g} and \ref{fig:wfn_map_1}. Scanning parameters: $V_{set}$ = 500 mV, $I_{set}$ = 200 pA, $V_{ac}$ = 10 mV, f = 773 Hz. a.u., arbitrary unit.}
\label{fig:wfn_map_2}
\end{figure}

High spatial resolution $dI/dV$ mapping of the 1T surface is shown in Fig.\ref{fig:wfn_map_1} and Fig.\ref{fig:wfn_map_2}. These figures show the rich evolution of the wave function distribution \cite{Wen2021}. Fig.\ref{fig:wfn_map_1}a shows the topography of the 1T layer over which the $dI/dV$ mapping shown in Fig.\ref{fig:wfn_map_1}b-f was measured. At energies about -300 to -400 meV we observe that the wave function is localized into three-balls structure within each CDW unit cell (Fig.\ref{fig:wfn_map_1}b). At energies close to the Fermi energy and below it we find that the wave function is localized on the CDW sites but its intensity varies across the CDW sites (Fig.\ref{fig:wfn_map_1}c,d and Fig.\ref{fig1}g). This variation in the intesity of the density of states may be caused by the inhomogenous charge transfer due to Se defects. A more homogenous wave function distribution centered on the CDW sites is observed at higher energies of about 200 meV (Fig.\ref{fig:wfn_map_1}e). The wave function localized on the CDW sites corresponds to the flat bands reminiscent of the Mott-Hubbard bands of 1T-TaS$_2$. On further increasing the energy, >300 meV, we find the wave function is delocalized corresponding to the conduction bands (Fig.\ref{fig:wfn_map_1}f). These features were also observed in a $dI/dV$ map measured in a different region with a different tip (Fig.\ref{fig:wfn_map_2}). These features in the wave function ditribution was qualitatively reproduced by DFT calculations, as shown in Fig.\ref{fig:DFT_maps}.

\FloatBarrier
\section{Density Functional Theory calculations} \label{fig:DFT_BS}
Figure \ref{fig:DFT_BS}a and b shows the theoretical result of 1T-TaS$_2$ band structure and density of state result with the SD structure. With the CDW reconstruction, 12 out of 13 Ta$^{4+}$ 5$d$-electrons SD clusters form molecular orbitals at lower valence bands or higher conduction bands. The one remained Ta 5$d$ electron at the center of SD makes a flat localized band near Fermi energy. From the onsite Coulomb interaction, this flat band is separated into an upper Hubbard band and a lower Hubbard band. As a result, the SD structure 1T-TaS$_2$ band structure shows Mott insulator character with the 200 meV Mott gap. 

Figure \ref{fig:DFT_BS}c and d shows the calculated bilayer 4H$_b$-TaS$_2$ band structure and density of state. The charge transfer from the 1T layer to the 1H layer is enough to shift the lower Hubbard band from the valence band to the conduction band. The electron-electron repulsion does not apply anymore, both bands sit at the conduction band with almost gapless. The Mott physics of 1T-TaS$_2$ has been removed in 4H$_b$ structure. Meanwhile, the band structure of the 1H part mainly comes from the Brillouin zone folding and hybridization with the 1T layer. The effect of $\sqrt{13}\times\sqrt{13}$ lattice reconstruction is much weaker compared with 1T layer transition.

Next from the calculated electronic structure, we simulated STM image of 4H$_b$-TaS$_2$ 1T layer surface. We integrated the charge distribution in the real space following to the specific energy region. The STM pattern from the flat bands, originated from the Mott feature of 1T layer, still sharply localized at the center of SD. On the other hand, the bands near Fermi energy except the flat bands are originated from the other 12 Ta atoms except center of Ta atom. They show slightly delocalized pattern than the flat bands.  

\begin{figure}[ht]
\centering
\includegraphics[scale=1]{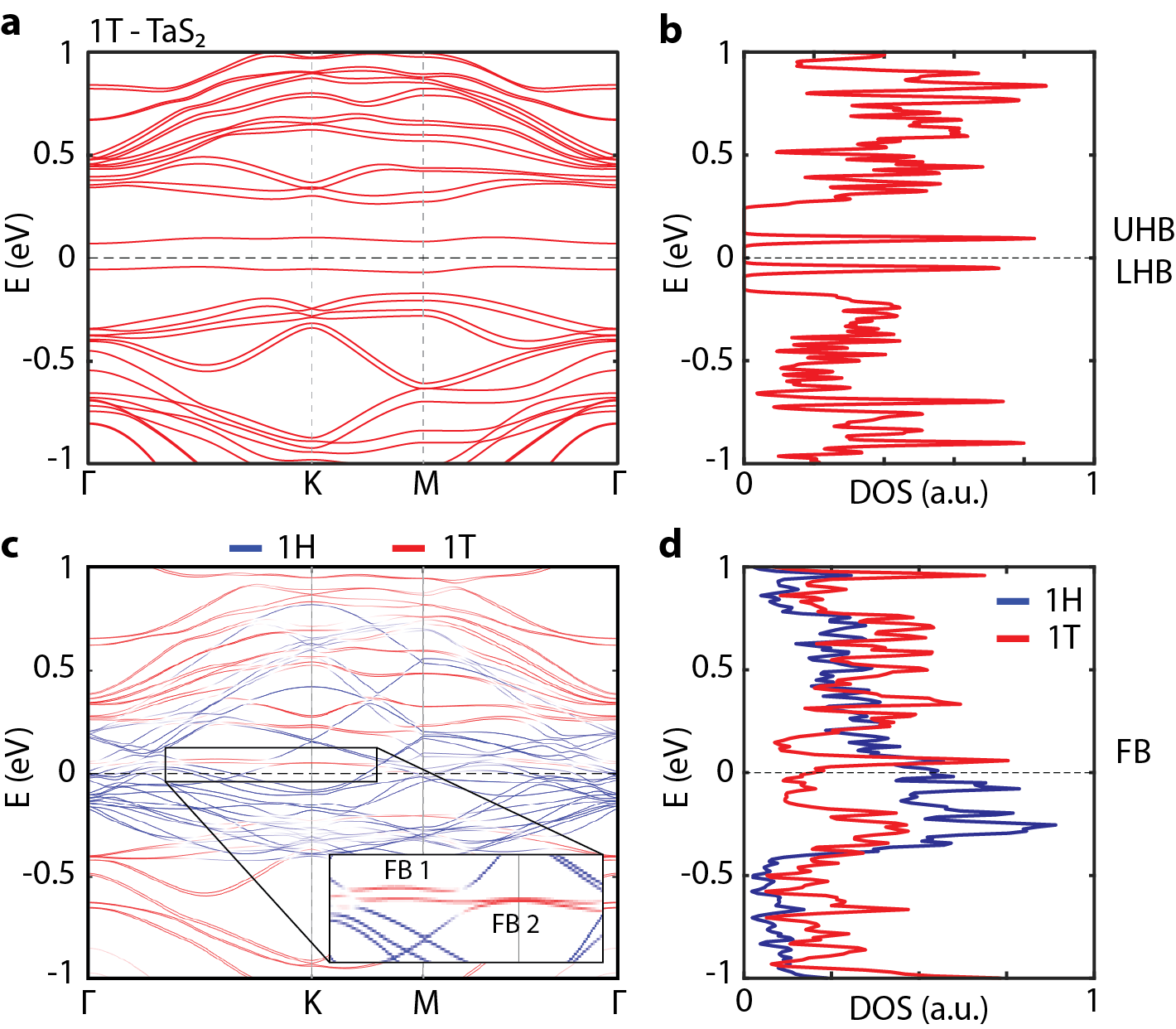}
\caption{\textbf{Calculated band structure of 1T-TaS$_2$ and bilayer 4Hb-TaS$_2$.} \textbf{a,} The calculated DFT band structure of a monolayer 1T-TaS$_2$ with a $\sqrt{13}\times\sqrt{13}$ CDW. \textbf{b,} The projected  density of states (DOS) for the monolayer 1T-TaS$_2$. The peaks in the DOS above and below Fermi energy corresponds to the upper (UHB) and the lower Hubbard bands (LHB). \textbf{c,} The calculated band structure of a bilayer 4Hb-TaS$_2$ with a CDW. The red and the blue color represents the weight of Ta atoms in the 1T and 1H layer, respectively. Inset shows a zoomed in image of the two flat bands (FB) just above the Fermi energy. \textbf{d} The projected density of states for the 1T and 1H layer Ta atoms for the bilayer 4Hb-TaS$_2$.}

\end{figure}

\begin{figure}[ht]
\centering
\includegraphics[scale=1]{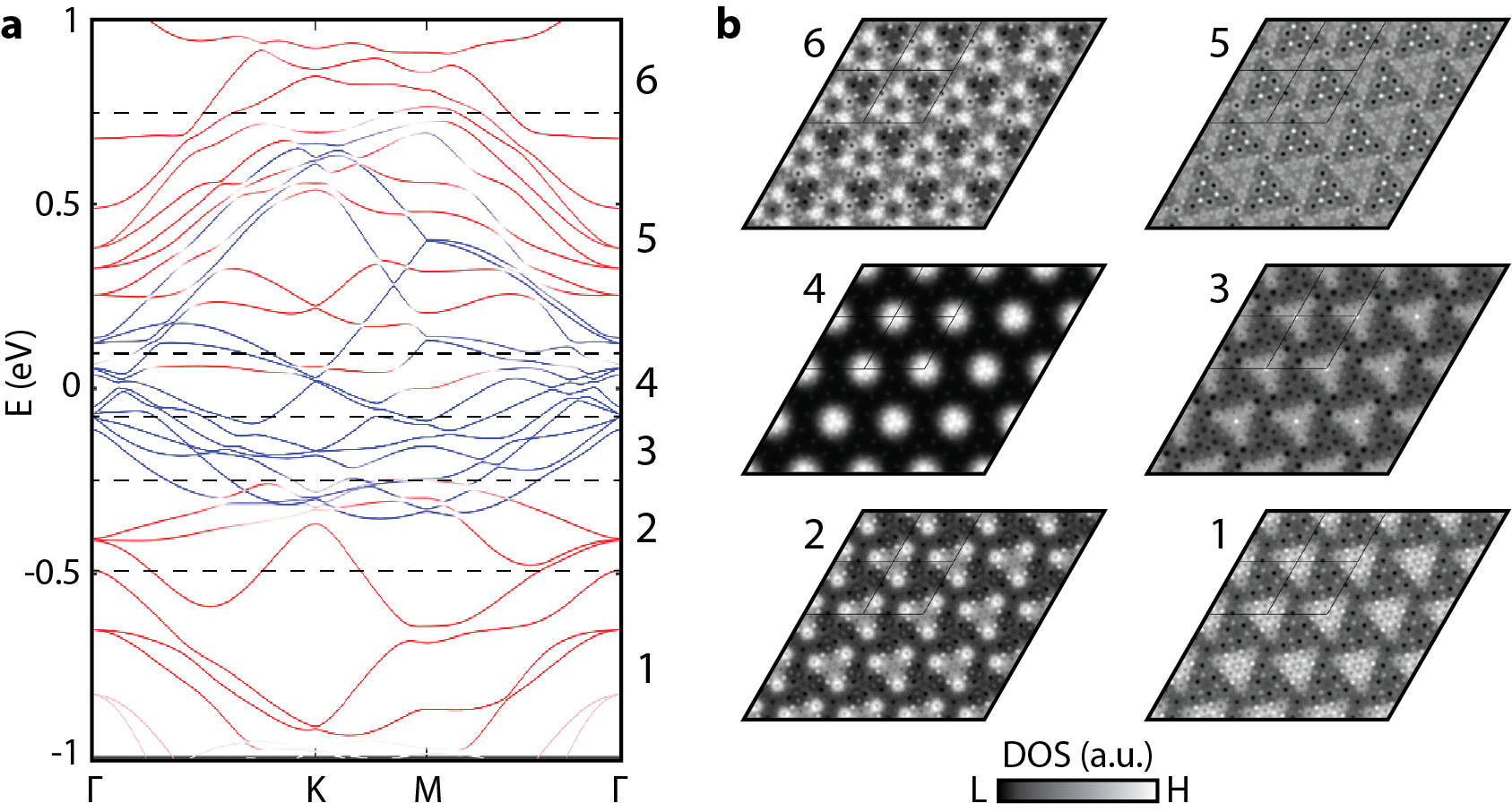}
\caption{\textbf{Calculated charge distribution on 1T layer.} \textbf{a,} The \textit{ab initio} calculated band structure of a bilayer 4Hb-TaS$_2$. The 1H and the 1T contributions are marked in blue and red, respectively. \textbf{b,} The calculated charge distribution on the 1T layer for different energy windows labeled from 6 to 1. The energy windows are marked in \textbf{a}.}
\label{fig:DFT_maps}
\end{figure}

\FloatBarrier

\section{QPI of flat bands} \label{fig:QPI}

We could not detect sharp QPI features in the I/dV maps taken on the 1T termination of 4Hb-TaS$_2$. The wave function distribution over the individual CDW sites appears fairly uniform which may be the cause for the absence of QPI. Still, we could identify faint features in Fourier space. They appeared within the energy window of 0-100 meV within which we find the DFBs. Two examples taken on two distinct 1T surfaces are given in Fig.\ref{fig:QPI1} and \ref{fig:QPI}. In them the Bragg peaks corresponding to the CDW periodicity appear clearly (atomic Bragg peaks are outside the Fourier space shown). Intriguingly, in both QPI maps within the energy range of the DFBs (72 meV cut is shown) we find arc-like features that replicate about the CDW Bragg peak. Such replication means that the wave function of the DFBs follows the periodicity of the CDW rather than the underlying atomic lattice. This suggests that the DFB wavefunction remains rather localized on the CDW sites. 

\begin{figure}[ht]
\centering
\includegraphics[scale=1]{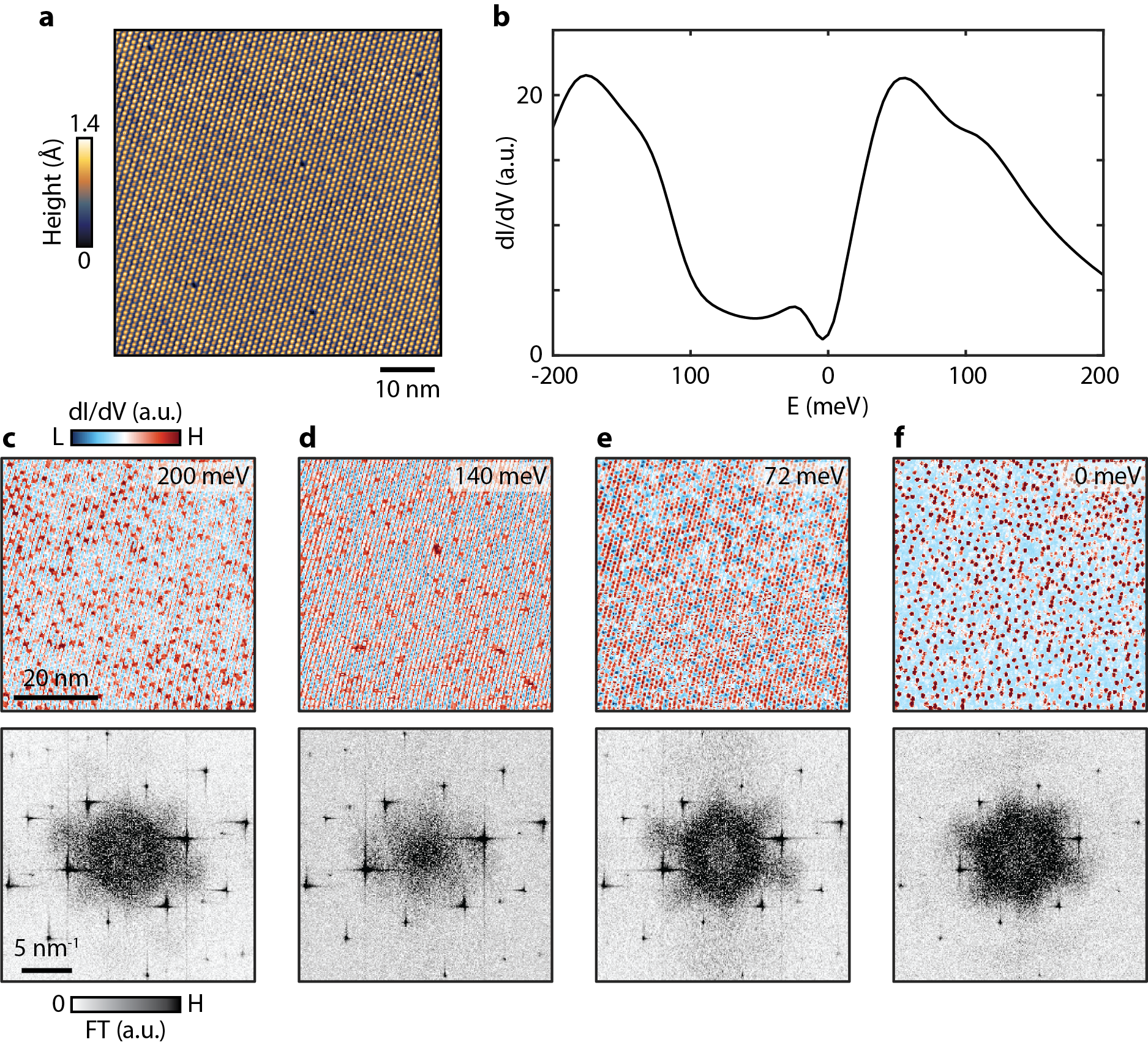}
\caption{\textbf{QPI.} \textbf{a,} Topographic image showing the CDW. \textbf{b,} dI/dV spectrum averaged over the entire surface shown in (a). \textbf{c-f,} dI/dV maps taken over the same field of view and their corresponding Fourier transform (top and bottom panels, respectively)} 
\end{figure} \label{fig:QPI1}

\begin{figure}[ht]
\centering
\includegraphics[scale=1]{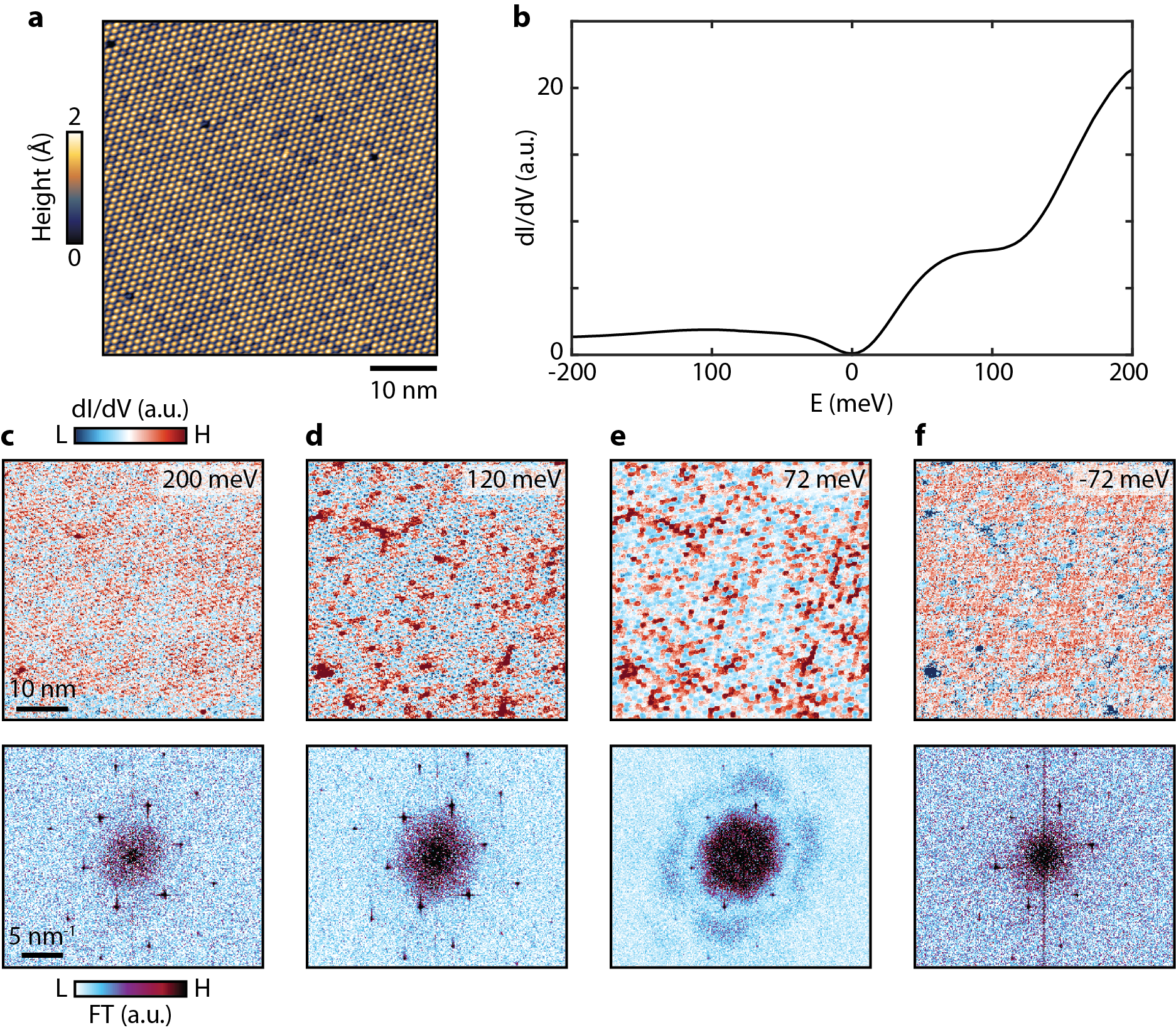}
\caption{\textbf{QPI.} \textbf{a,} Topographic image showing the CDW. \textbf{b,} dI/dV spectrum averaged over the entire surface shown in (a). \textbf{c-f,} dI/dV maps taken over the same field of view and their corresponding Fourier transform (top and bottom panels, respectively)} 
\label{fig:QPI2}
\end{figure}

\FloatBarrier


\section{Analysis of `dim' sites in topography} \label{fig:topo_dim}
A typical topography of the 1T layer is shown in Fig.\ref{fig:topo_dim}a (same as Fig.1b). We identify two different types of defects in the CDW superlattice: (i) `dim', as shown in Fig.\ref{fig:topo_dim}b and (ii) `vacancy', as shown in Fig.\ref{fig:topo_dim}c. Their reduced height profile readily identifies the `dim' sites compared to the regular (`bright') CDW sites. The various CDW sites were identified by appropriately using an intensity threshold and `regionprops' function in MATLAB. The identified 'dim' sites are overlaid on the topography as shown in Fig.\ref{fig:topo_dim}d. The 'dim' sites constitute $\sim 12\%$ of the entire CDW sites in this topography. We surmise that the 'vacancy' in the CDW lattice is due to Ta vacancy and the 'dim' sites are due to either S vacancy or Se substitution \cite{Liu2021}.

\begin{figure}[ht]
\centering
\includegraphics[width=\linewidth]{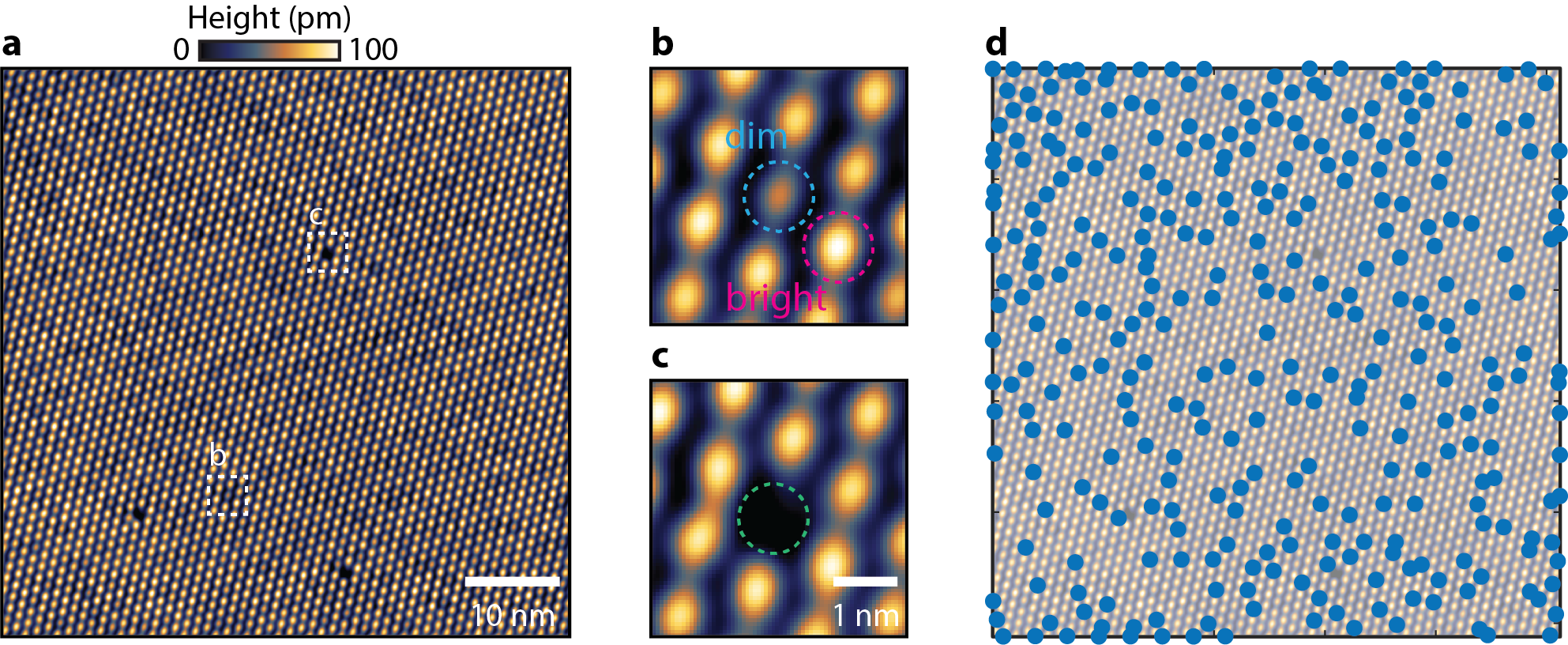}
\caption{\textbf{Analysis of 'dim' sites in the CDW lattice.} \textbf{a,} Topography of the 1T layer showing commensurate CDW. \textbf{b,} Zoom in of \textbf{a} highlighting the 'dim' and 'bright' CDW sites. \textbf{c,} Zoom in of \textbf{a} shown a vacancy-like defect in the CDW lattice. \textbf{d,} The 'dim' CDW sites, amounting to $\sim 12$\% of the total CDW sites, are marked by blue circles overlaid on top of the topography. Scanning parameters: $V_{set}$ = 200 mV, $I_{set}$ = 50 pA.}

\end{figure}

\FloatBarrier

\section{Kondo resonance}
\begin{figure}[ht]
\centering
\includegraphics[scale=1]{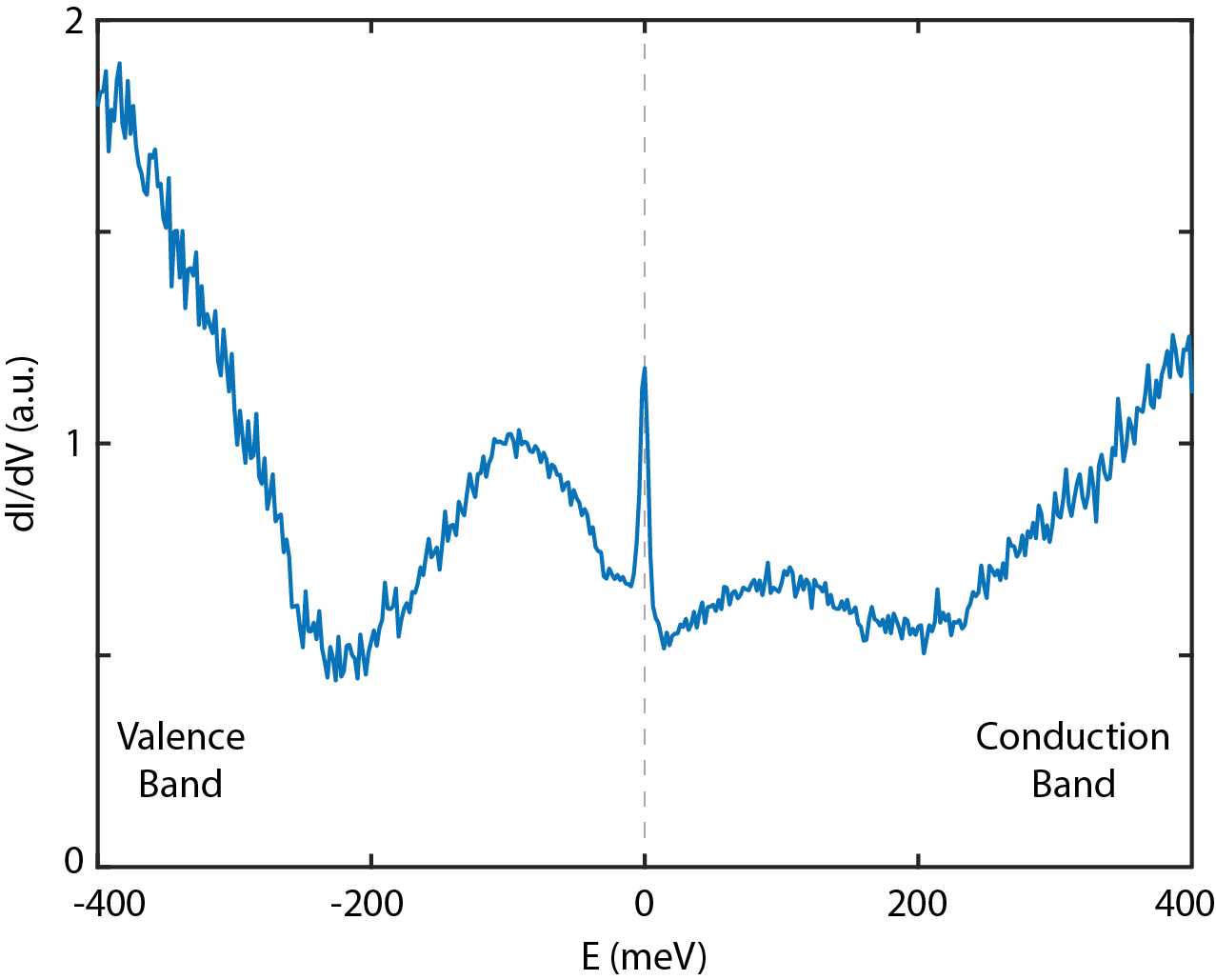}
\caption{\textbf{Large energy window spectroscopy of the Kondo resonance.} The quasi-particle peaks at finite energies, $\sim 100$ meV above and below Fermi energy, correspond to the upper and lower Hubbard band, respectively. Scanning parameters: $V_{set}$ = 400 mV, $I_{set}$ = 100 pA, $V_{ac}$ = 3 mV, f = 773 Hz. a.u., arbitrary unit.}
\label{fig:ZBP_1}
\end{figure}

A representative large energy window spectrum of the Kondo resonance is shown in Fig.\ref{fig:ZBP_1}. The spectrum was measured on a `dim' CDW site of the 1T layer (same as Fig. 2a). The conduction and the valence band emerge at high energies $\sim 200$ meV above and below the Fermi energy (see also Fig.\ref{fig:DFT_BS}a and b). The upper and lower Hubbard bands appear as quasi-particle peaks $\sim 100$ meV away from the Fermi energy. The sharp zero-bias peak is identified as the Kondo resonance. 

\FloatBarrier

\section{Fitting analysis of Kondo resonance}
\begin{figure}[ht]
\centering
\includegraphics[scale=1]{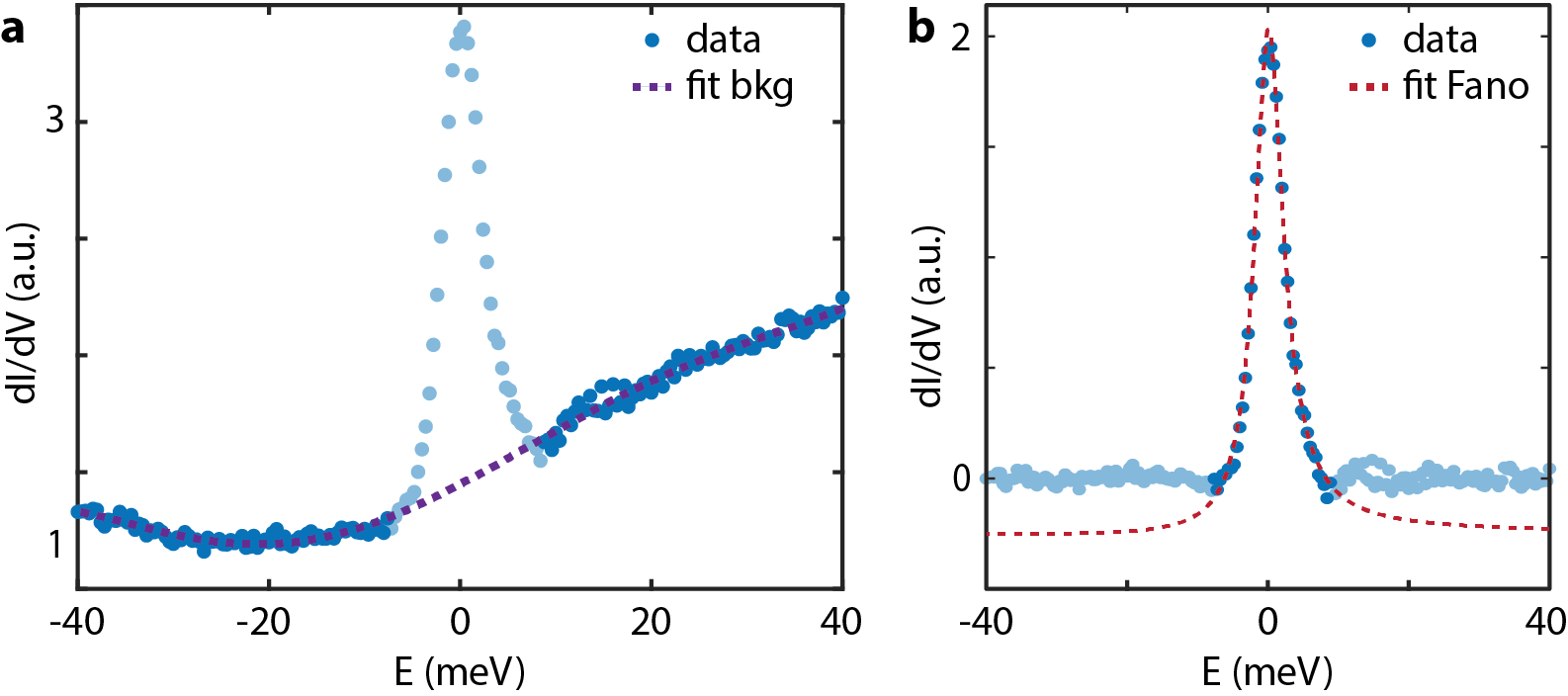}
\caption{\textbf{Fano fit to the Kondo resonance.} \textbf{a,} dI/dV profile of a Kondo resonance. The background (bkg) was removed by fitting the data excluding the zero-bias peak (dark blue) to a smooth polynomial. \textbf{b,} The background subtracted Kondo resonance was fit to a Fano line shape $\propto (q+\epsilon)^2/(1+\epsilon^2)$, where $q$ is the form factor and $\epsilon$ is the normalized energy $\epsilon = (E-E_0)/\Gamma$ with $E_0$ as the resonance position and $\Gamma$ as the half width at half maximum. The Fano fit parameters are: q = 22(10), $\Gamma = 2.5(1)$ meV, $E_0 = 0.0(1)$ meV. Scanning parameters: $V_{set}$ = 40 mV, $I_{set}$ = 100 pA, $V_{ac}$ = 1 mV, f = 773 Hz. a.u., arbitrary unit.}
\label{fig:fano_fit}
\end{figure}

High energy resolution dI/dV profile of a Kondo resonance is shown in Fig.\ref{fig:fano_fit}a. A smooth polynomial background was subtracted from the dI/dV profile. A fifth-order polynomial was fit to the dark blue circles in Fig.\ref{fig:fano_fit}a, marked by the dashed purple line. The subtracted profile is shown in Fig.\ref{fig:fano_fit}b (also in Fig.2b). The Kondo resonance is often fit to a Fano line shape in STM measurements. The Fano line shape is given by $f(\epsilon) = \propto (q+\epsilon)^2/(1+\epsilon^2)$, where $q$ is the form factor and $\epsilon$ is the normalized energy $\epsilon = (E-E_0)/\Gamma$ with $E_0$ as the resonance position and $\Gamma$ as the half-width at half-maximum. The dI/dV profile (dark blue circles) in Fig.\ref{fig:fano_fit}b is fit to $A*f(\epsilon)+B$, where $A$ and $B$ are additional fitting parameters. The Fano fit parameters are: q = 22(10), $\Gamma = 2.5(1)$ meV, $E_0 = 0.0(1)$ meV. The high value of $q$ justifies using a simpler Lorentzian line shape to capture the width of the Kondo resonance. The half-width at half-maximum from both the Fano fit ($2.5(1)$ meV) and the Lorentzian fit ($2.6(2)$ meV) are very similar.

\FloatBarrier

\section{Magnetic field response of Kondo resonance}
\begin{figure}[ht]
\centering
\includegraphics[scale=1]{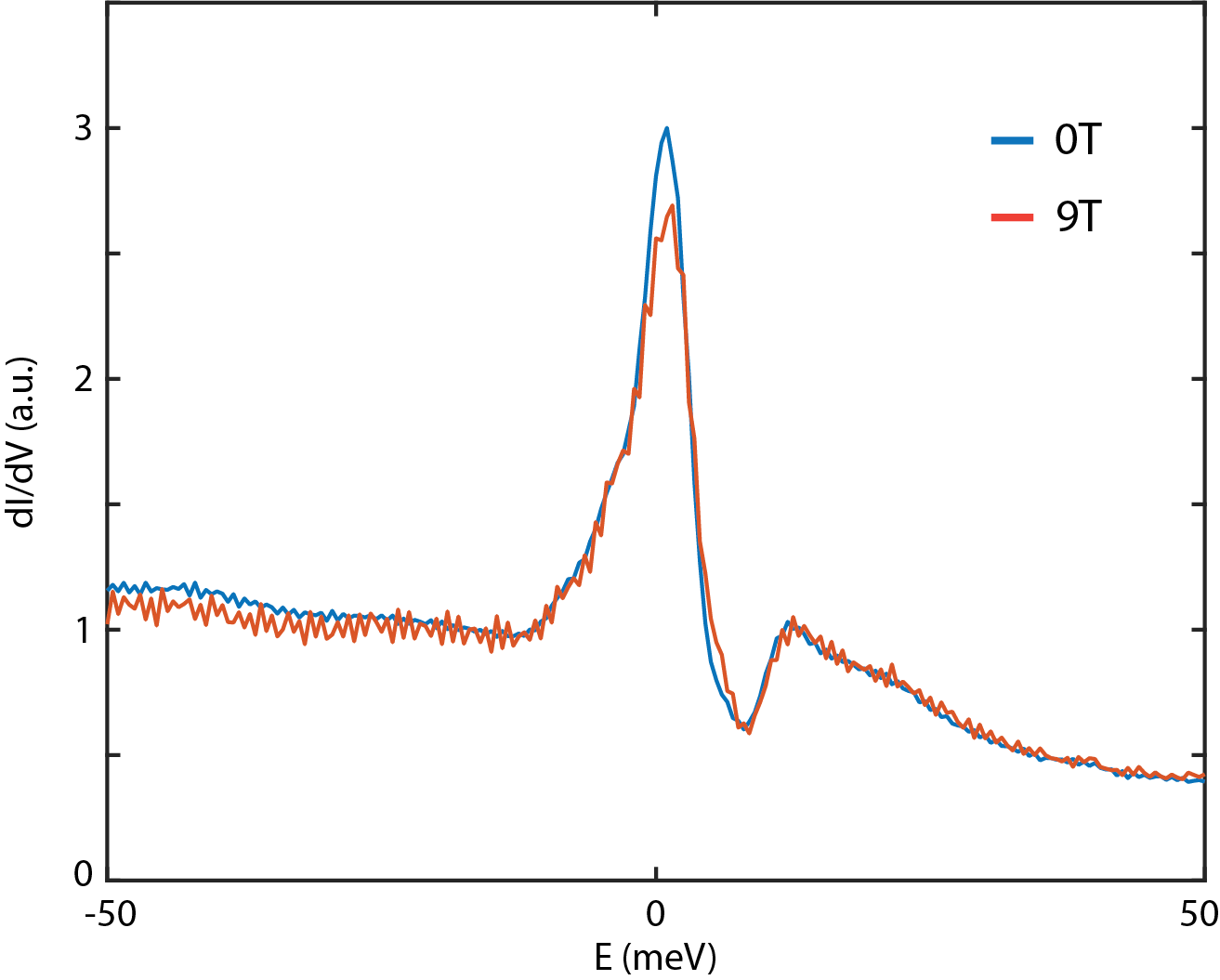}
\caption{\textbf{Magnetic field response of the Kondo resonance.} The high energy resolution dI/dV profile of the zero-bias peak remains unchanged at the highest out-of-plane magnetic field of 9T. Scanning parameters: $V_{set}$ = 50 mV, $I_{set}$ = 400 pA, $V_{ac}$ = 1 mV, f = 433 Hz. a.u., arbitrary unit.}
\label{fig:ZBP_B}
\end{figure}

We did not observe any significant change in the Kondo resonance up to the maximum out-of-plane magnetic field of 9T possible to apply in our system (Fig.\ref{fig:ZBP_B}). This lack of magnetic field response could be due to that Zeeman energy at 9T ($E_Z = g\mu_B B \sim 1$ meV) is not enough to split the Kondo resonance ($\sim 2.5$ meV) \cite{Zhang2020}. 

\FloatBarrier


\section{Spectroscopic maps of Kondo resonance} \label{fig:ZBP_MAPS}
\begin{figure}[ht]
\centering
\includegraphics[scale=1]{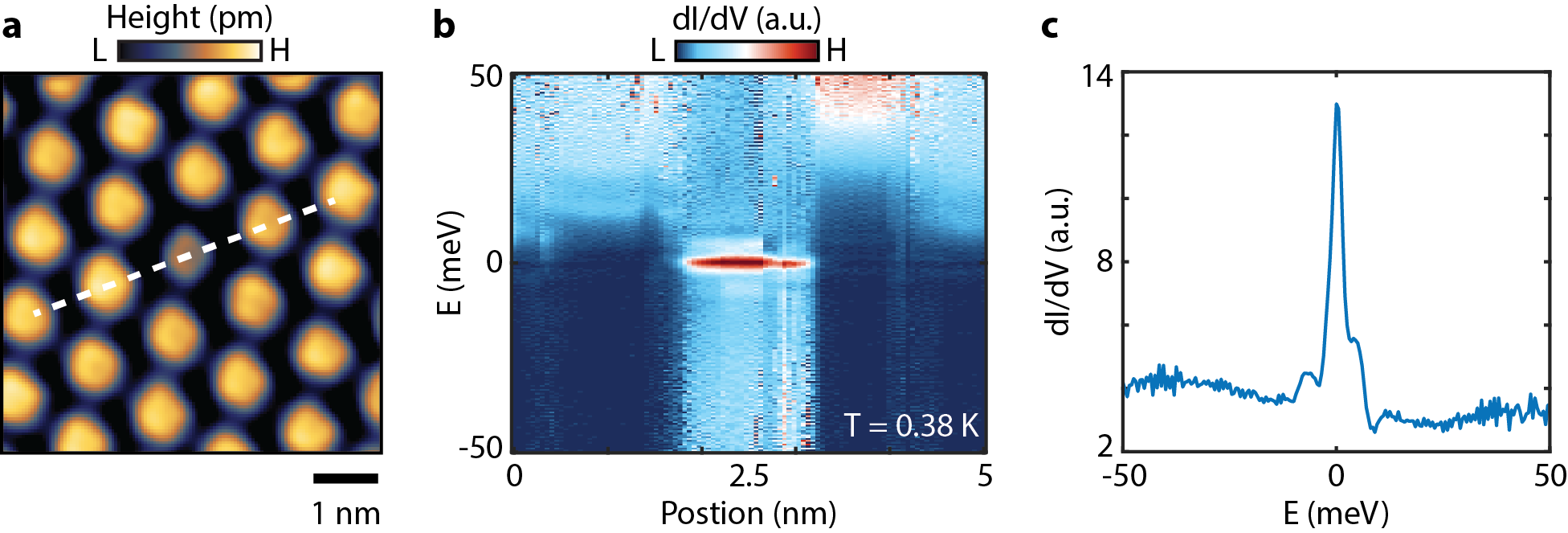}
\caption{\textbf{High energy resolution dI/dV mapping of the Kondo resonance.} \textbf{a,} Topography of the 1T layer with a single 'dim' CDW site in the center. \textbf{b,} dI/dV map across the 'dim' site (marked by white dashed line in \textbf{a}) shows a sharp zero-bias peak at the 'dim' site identified as the Kondo resonance. \textbf{c,} dI/dV profile of the Kondo resonance extracted from \textbf{b}. Scanning parameters: $V_{set}$ = 50 mV, $I_{set}$ = 200 pA, $V_{ac}$ = 1 mV, f = 433 Hz. a.u., arbitrary unit.}
\label{fig:ZBP_LC_2}
\end{figure}

\begin{figure}[ht]
\centering
\includegraphics[width=\linewidth]{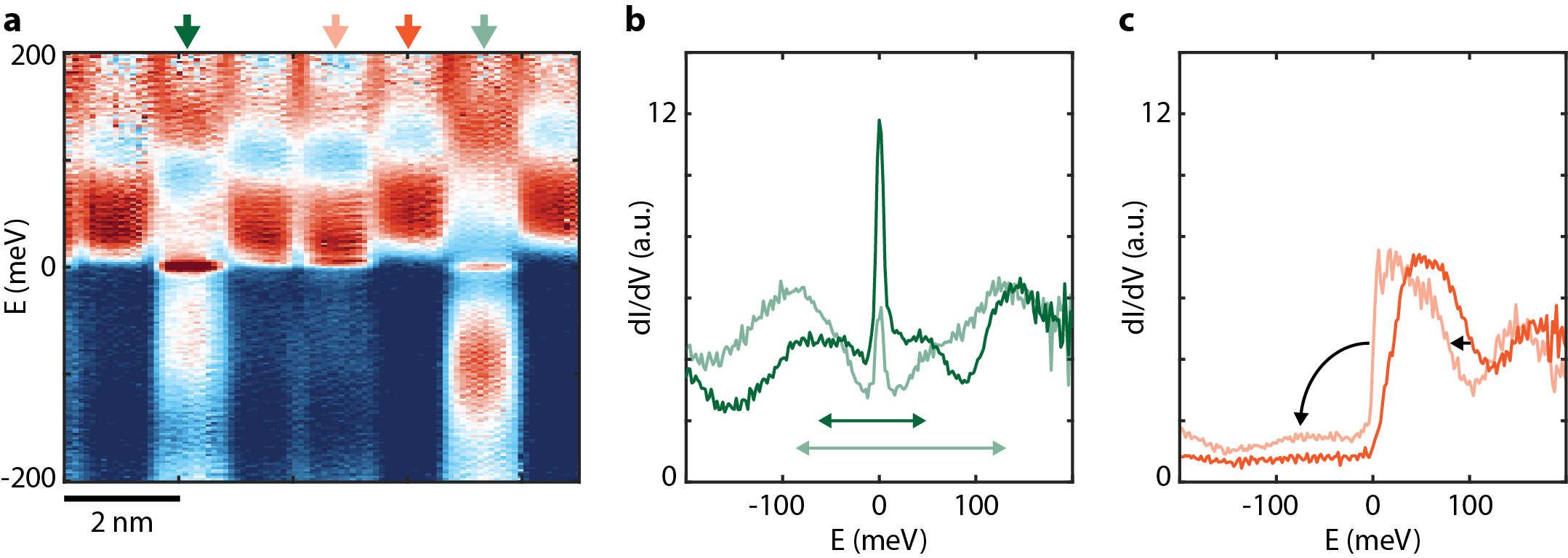}
\caption{\textbf{1D dI/dV map showing Kondo resonance.} \textbf{a,} Spectroscopic map across CDW of the 1T layer showing the Kondo resonance. \textbf{b,} The dI/dV profiles extracted from \textbf{a} (marked by arrows) shows the diversity of spectrum on the 1T layer. Scanning parameters: $V_{set}$ = 200 mV, $I_{set}$ = 200 pA, $V_{ac}$ = 5 mV, f = 433 Hz. a.u., arbitrary unit.}
\label{fig:ZBP_LC_1}
\end{figure}

\begin{figure}[ht]
\centering
\includegraphics[scale=1]{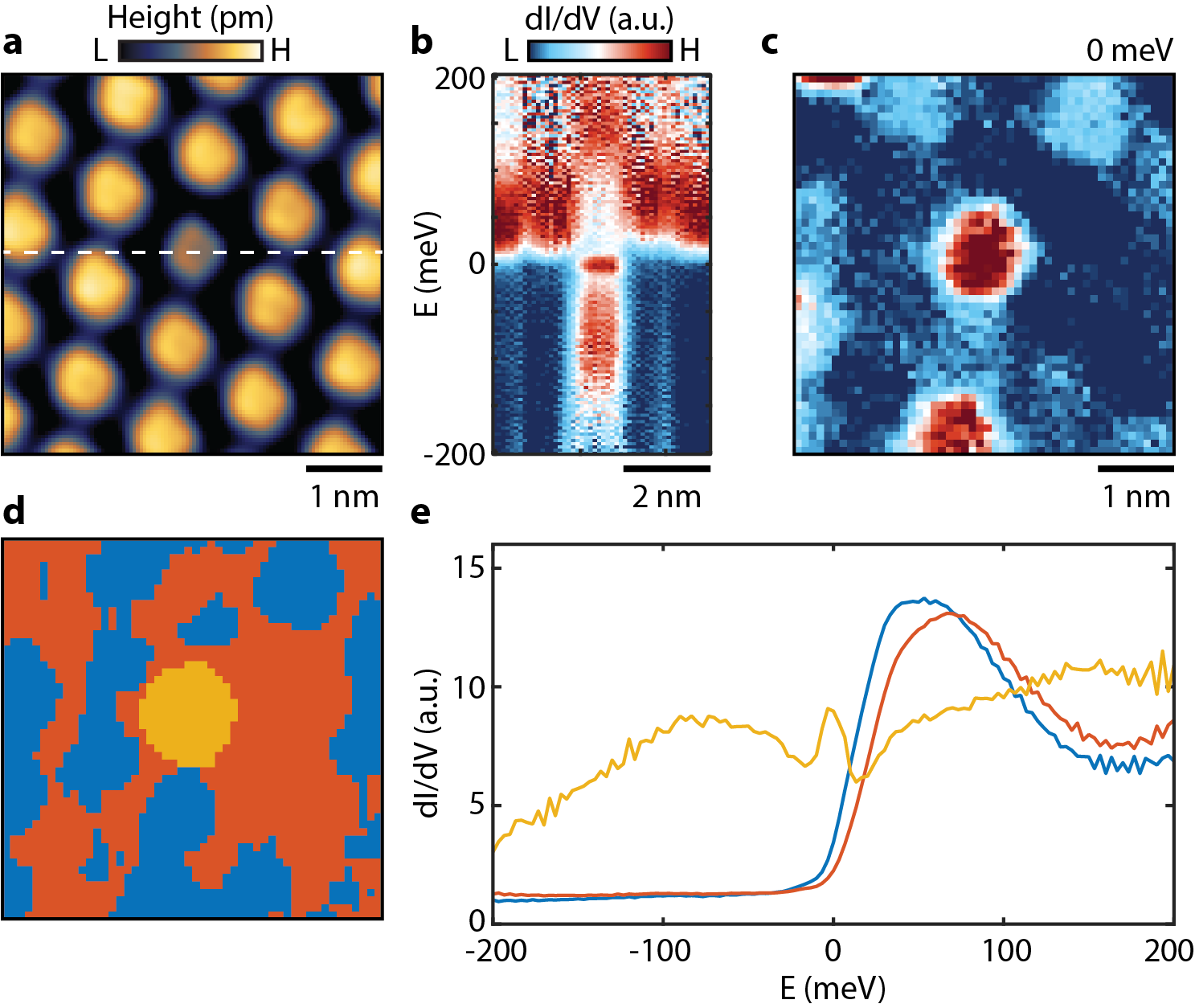}
\caption{\textbf{Cluster analysis of the spectroscopic map on the 1T layer.} \textbf{a,} Topography of the 1T layer in the same region as \ref{fig:ZBP_LC_2}. \textbf{b,} 1D dI/dV map across the 'dim' CDW site exhibiting Kondo resonance. \textbf{c,} Zero-bias energy slcie of the 2D dI/dV map measured on \textbf{a}. \textbf{d,} Cluster analysis of the 2D dI/dV map measured on \textbf{a}. Similar spectra are represented with the same color. \textbf{e,} The cluster analysis yields two distinct spectra: (i) zero-bias peak (yellow) and (ii) quasi-particle peak at $\sim 50$ meV (blue and orange). The Kondo resonance (yellow) is localized on the 'dim' site only. Scanning parameters: $V_{set}$ = 50 mV, $I_{set}$ = 200 pA, $V_{ac}$ = 1 mV, f = 433 Hz. a.u., arbitrary unit.}
\label{fig:ZBP_map}
\end{figure}

1D and 2D dI/dV maps similar to Fig.2c are shown in Fig.\ref{fig:ZBP_LC_1} - \ref{fig:ZBP_map}. Fig.\ref{fig:ZBP_LC_1}a shows the dI/dV measurement across several 'bright' and a couple of 'dim' CDW sites. The 'bright' CDW sites typically show a quasi-particle peak at about 70 meV above the Fermi energy as shown in Fig.\ref{fig:ZBP_LC_1}b (yellow). However, sometimes we also see a spectrum where the quasi-particle peak is closer to the Fermi energy along with a sharp drop in the density of states across the Fermi energy (red). The 'dim' sites exhibit a robust zero-bias peak (or Kondo resonance) (blue). The Kondo resonance is localized to the 'dim' CDW site, as also shown in Fig.\ref{fig:ZBP_LC_2}. 

2D mapping of the 'dim' CDW sites reveals the entire spatial distribution of the Kondo resonance (Fig.\ref{fig:ZBP_map}). The topography of the 1T CDW lattice with a single 'dim' site in the center is shown in Fig.\ref{fig:ZBP_map}a. A 1D dI/dV map across the 'dim' site, marked by a white dashed line, is shown in Fig.\ref{fig:ZBP_map}b. The dI/dV map measured on Fig.\ref{fig:ZBP_map}a at the Fermi energy is shown in Fig.\ref{fig:ZBP_map}c. The 'dim' site shown a high density of states corresponding to the Kondo resonance. To better visualize the different types of spectra, we cluster all the spectra measured in this region. The spectra are clustered into three different types denoted by different colors. Fig.\ref{fig:ZBP_map}d shows the spatial region of the different spectra, and Fig.\ref{fig:ZBP_map}e shows the averaged spectra of each type.

\begin{figure}[ht]
\centering
\includegraphics[scale=1]{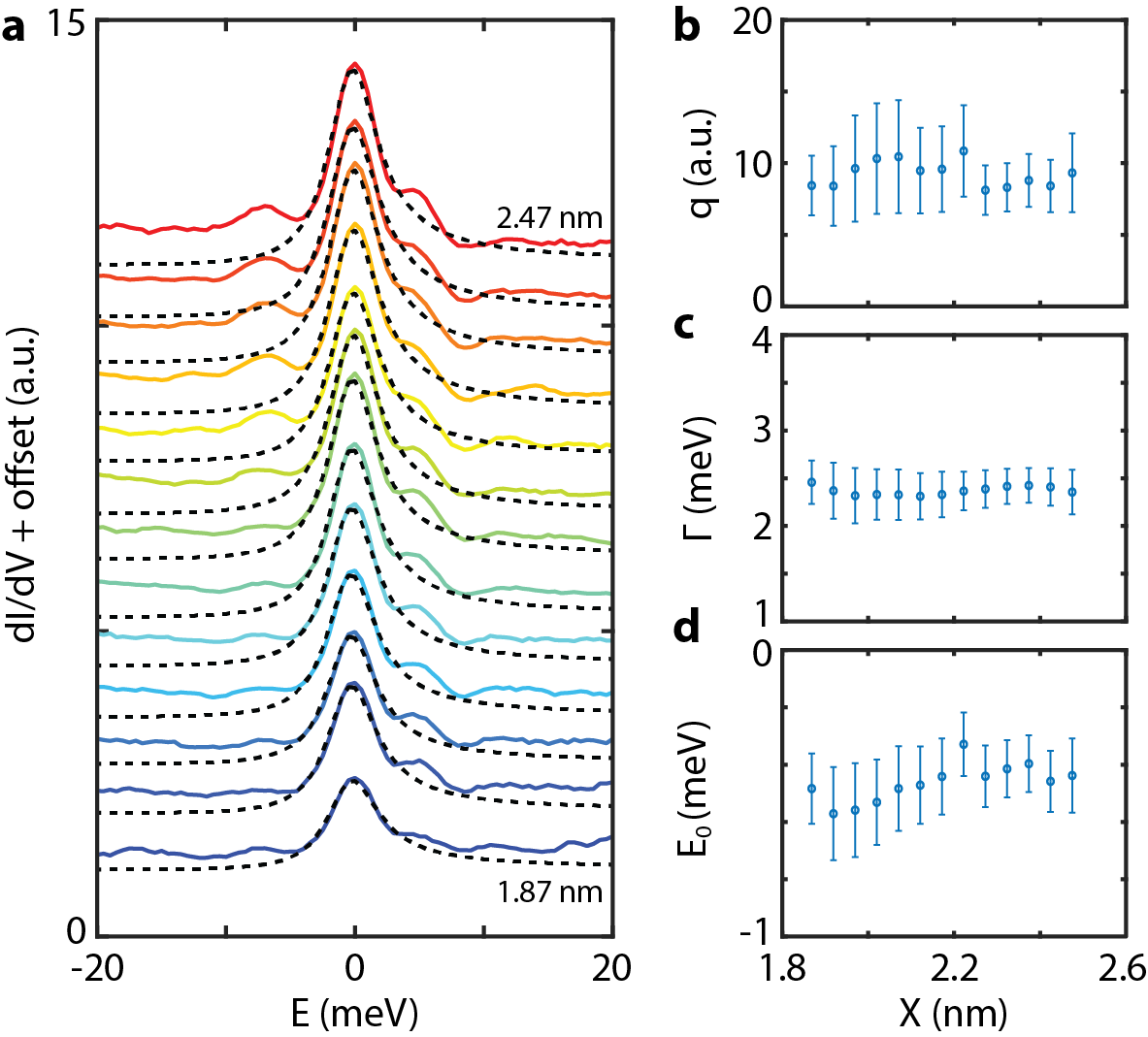}
\caption{\textbf{Spatial resolved Fano fits.} \textbf{a,} dI/dV profiles of the zero-bias peak measured on the edge (at X = 1.87 nm of \ref{fig:ZBP_LC_2}\textbf{b}) towards the center of the 'dim' CDW (at X = 2.47 nm) and the corresponding Fano fits. The spectra were smoothened by a Savitzky-Golay filter for robust fitting. \textbf{b-d,} The fitting parameters of the Fano line shape: quality factor (q), peak half width ($\Gamma$), and peak position ($E_0$), respectively.}
\label{fig:ZBP_LC_Fano_fit}
\end{figure}

\FloatBarrier

\section{Hysteresis}
\begin{figure}[ht]
\centering
\includegraphics[scale=1]{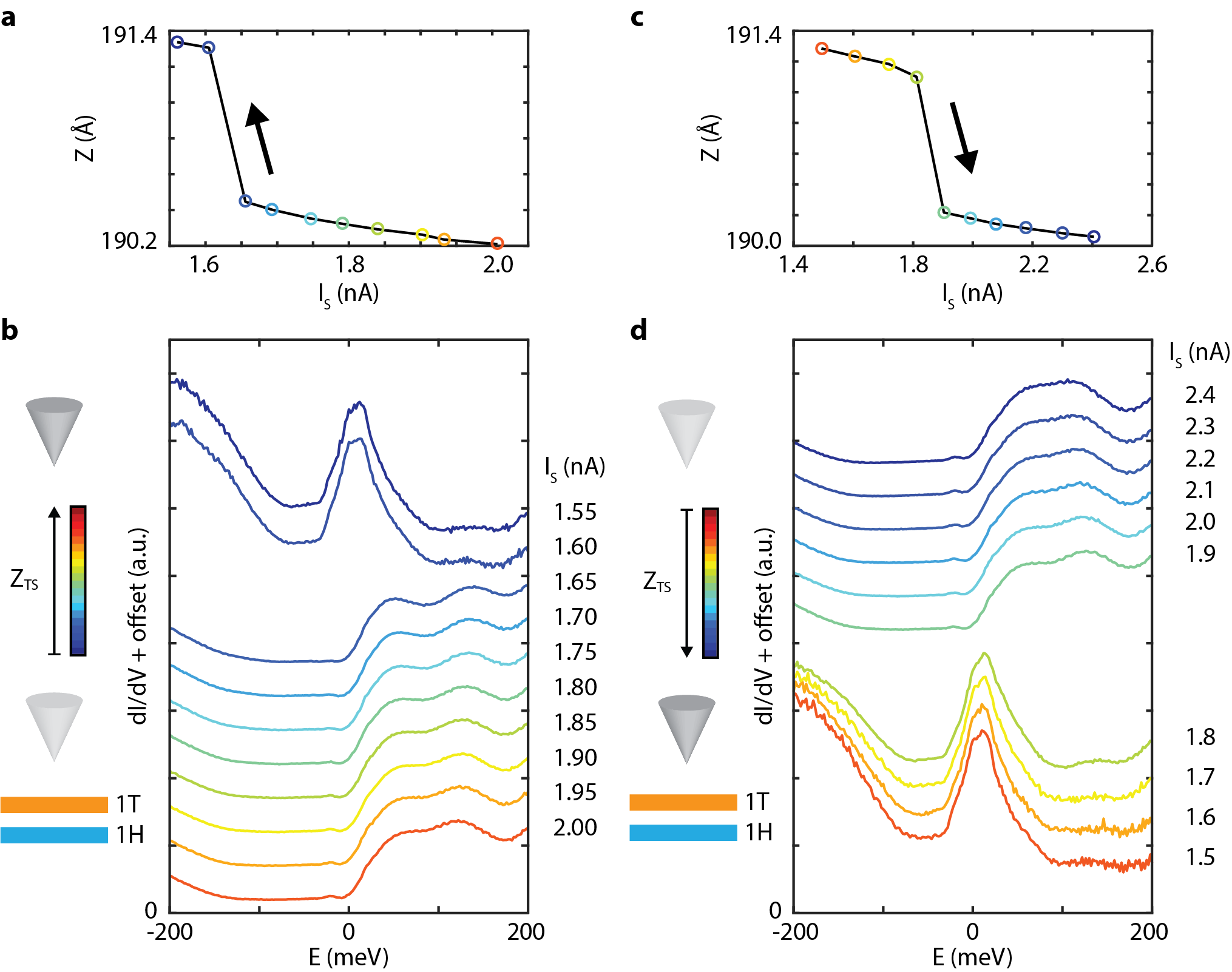}
\caption{\textbf{Hysteresis.} \textbf{a(c),} Current set poin condition at which the step in tip position occurs on retracting (approaching) sweeps. \textbf{b(d),} corresponding dI/dV spectra that change abruptly across the transition point on retracting (approaching) sweeps.}
\label{fig:Hysteresis}
\end{figure}

By taking fine steps in tip position about the phase transition point (marked both by an abrupt change of spectrum in Fig.\ref{fig:Hysteresis}a,c as well as by a resulting discontinuous step in tip height  in Fig.\ref{fig:Hysteresis}b,d) we find a hysteresis in the tip distance from sample at which the transition takes place between approaching and retracting sweep directions. 

\FloatBarrier
\section{Noise}
\begin{figure}[ht]
\centering
\includegraphics[width=\linewidth]{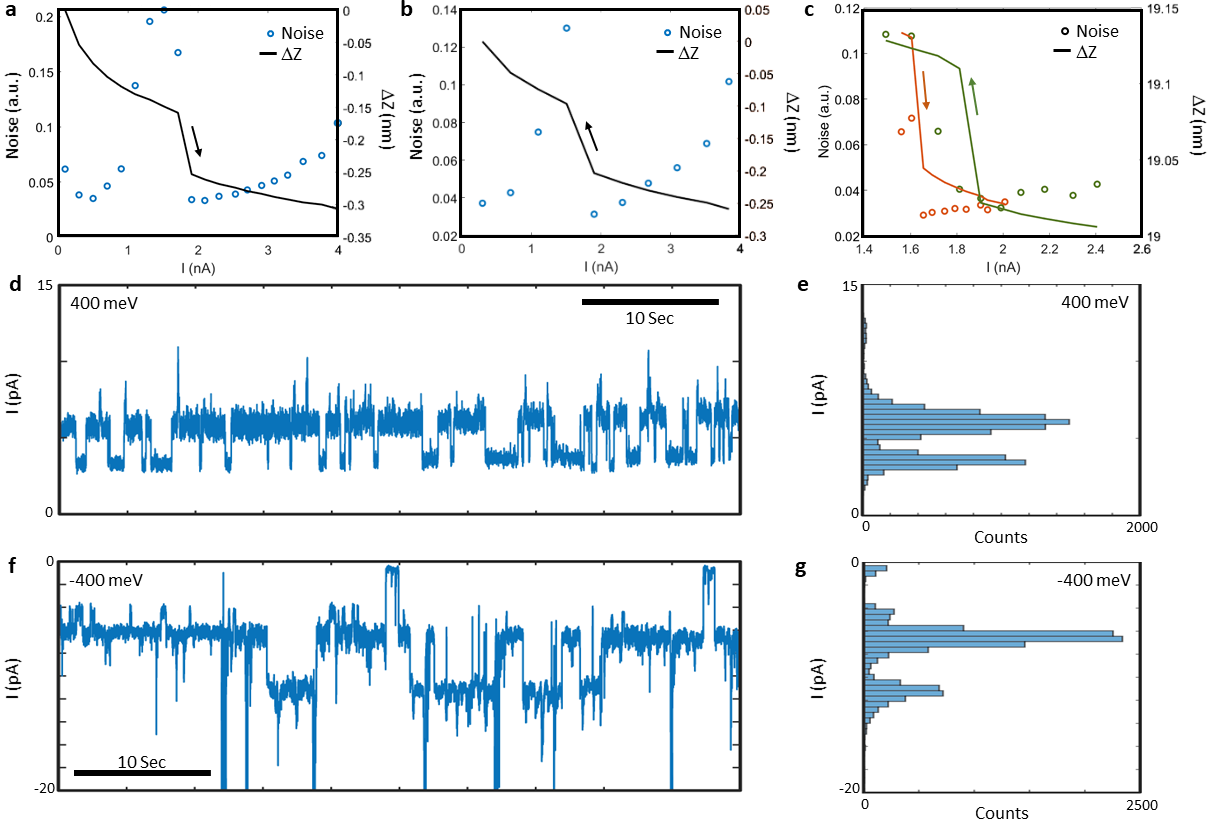}
\caption{\textbf{Noise characteristic of dim CDW sites.} \textbf{a(b),} tip height above sample in solid line and standard deviation among 10 measured spectra averaged over the whole measured energy window on the approaching (retracting) height sweep varied by modifying the requested set point current, $I_s$. \textbf{c,} Similar height and noise profiles for both retracting and approaching sweeps (green and orange, respectively) taken over fine steps about the transition point. \textbf{d(f)} Time trace of measured tunneling current taken with the feed back loop open over a dim CDW site at 400 meV (-400 meV). \textbf{e(g)} Hystogram of the time trace in d(f) showing clear bi-stability in the tunneling current characteristic of telegraphic noise.} 
\label{fig:noise}
\end{figure}

The dI/dV spectra taken about the transition point seem noisier. We quantify this by calculating the standard deviation of the 10 consecutive measurements we take at every energy point (typically we show their mean). In Fig.\ref{fig3}a and in Fig.\ref{fig:noise}a-c we show the mean value of that calculated standard deviation over all energies measured. In rough position sweeps, shown in Fig.\ref{fig:noise}a,b, we find that the noise increases sharply about the transition point on the ZBCP side of the spectrum (regardless of sweep direction). When we take finer steps about the transition, where we resolve the hysteresis loop, we also find slight asymmetry between approaching and retracting sweeps. On the approaching sweep (orange) the noise rises and drops sharply at the transition point. On the retracting sweep (green) we see a gradual onset of the noise already before it takes place and drops gradually across it. We note that all our measurements are taken with at a very low frequency of about 1 KHz. Therefore, the noise we find has low frequency characteristics. Intriguingly on some of the dim CDW sites we also find a bi-stable noise pattern that switches between two values as shown in the time traces in Fig.\ref{fig:noise}d,f. They show the tunneling current taken with the feedback loop open, thus reflecting changes in the electronic system of 1T-TaS$_2$ rather than of the tip (we do not detect such behavior away from dim CDW sites). Their histograms, given in Fig.\ref{fig:noise}e,g, respectively, clearly show bi-stability characteristic of telegraphic noise.

\FloatBarrier

\section{Tip manipulation of the Kondo resonance}
\begin{figure}[ht]
\centering
\includegraphics[scale=1]{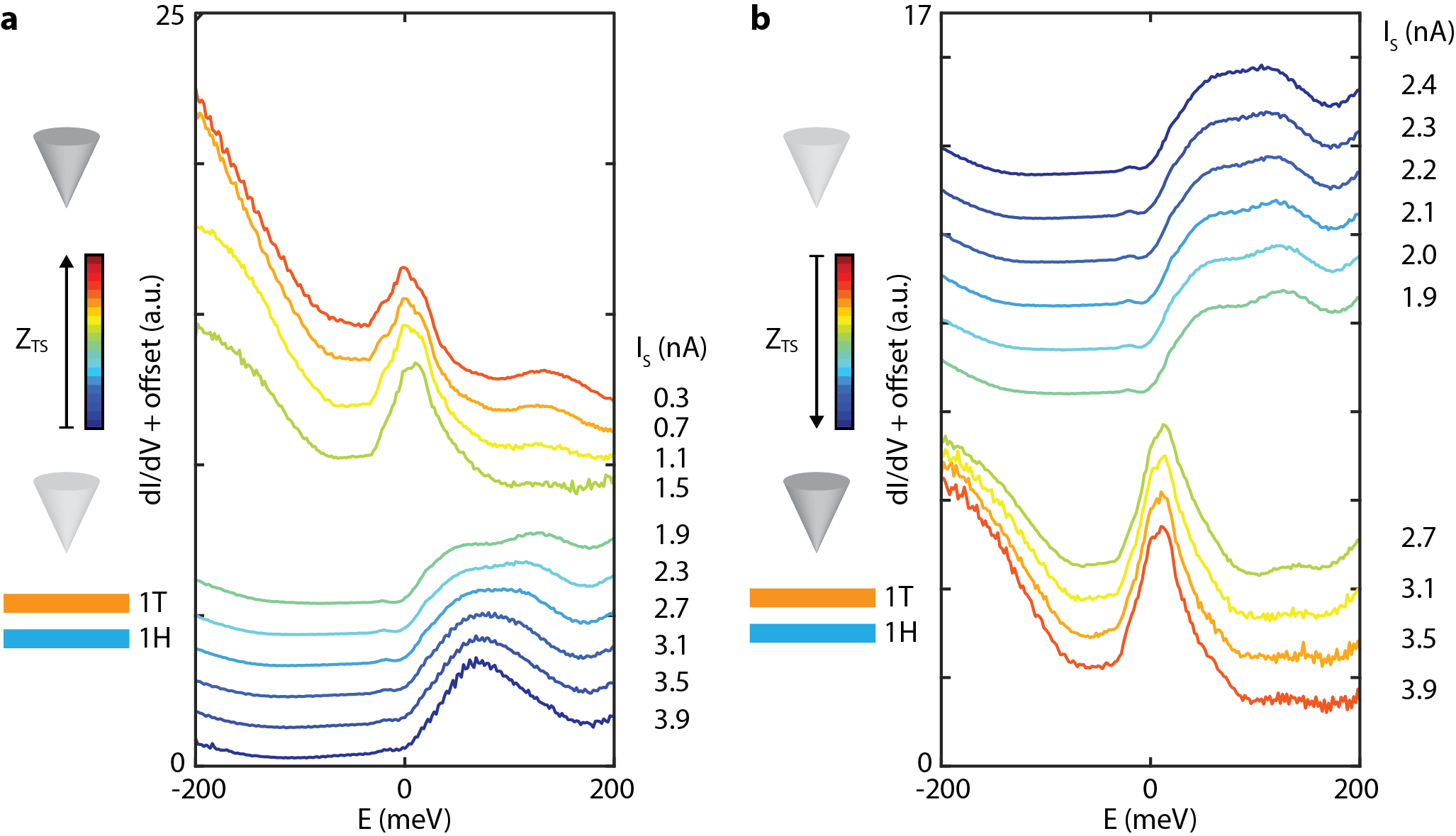}
\caption{\textbf{Reversible tip manipulation of the Kondo resonance.} \textbf{a,} dI/dV profiles measured on a 'dim' CDW site (same site as in Fig.3\textbf{c}) while retracting the tip away from the sample. As the tip-sample distance increases, the spectrum abruptly transitions from peaks at finite energy to a zero-bias peak. Scanning parameters: $V_{set}$ = 200 mV, $I_{set}$ = 3900 pA to 300 pA in steps of 400 pA, $V_{ac}$ = 5 mV, f = 433 Hz. a.u., arbitrary unit. \textbf{b,} Repeating the measurement shown in Fig.3\textbf{c} reproduces the abrupt transition of the Kondo peak showcasing reversible tip manipulation. Scanning parameters: $V_{set}$ = 200 mV, $I_{set}$ = 1500 pA to 2400 pA in steps of 100 pA, $V_{ac}$ = 5 mV, f = 433 Hz.}
\label{fig:ZBP_IZ}
\end{figure}

The spectroscopy measurements presented in Fig.3 are reversible and reproducible. Similar measurements are presented in Fig.\ref{fig:ZBP_IZ} for both tip direction and step heights. The tip-sample distance ($Z_{TS}$) is changed by changing the current set-point ($I_set$). Increasing the $I_{set}$ decreases the tip-sample distance and vice-versa. Fig.\ref{fig:ZBP_IZ}a shows the reverse measurement of Fig.3c, i.e., we start with the tip very close to the sample (blue) and gradually withdraw the tip away from the sample (red). We see the reverse phenomenology. When the tip is near the sample, a single quasi-particle peak is observed at $\sim 70$ meV. This peak splits as the tip-sample distance decreases. Eventually, the spectrum transitions back to the Kondo resonance. Repeating this measurement in the opposite direction with much smaller increments of tip-sample distance leads to the same result as shown in Fig.\ref{fig:ZBP_IZ}b.

\section{Tip manipulation of flat bands}
\begin{figure}[ht]
\centering
\includegraphics[scale=1]{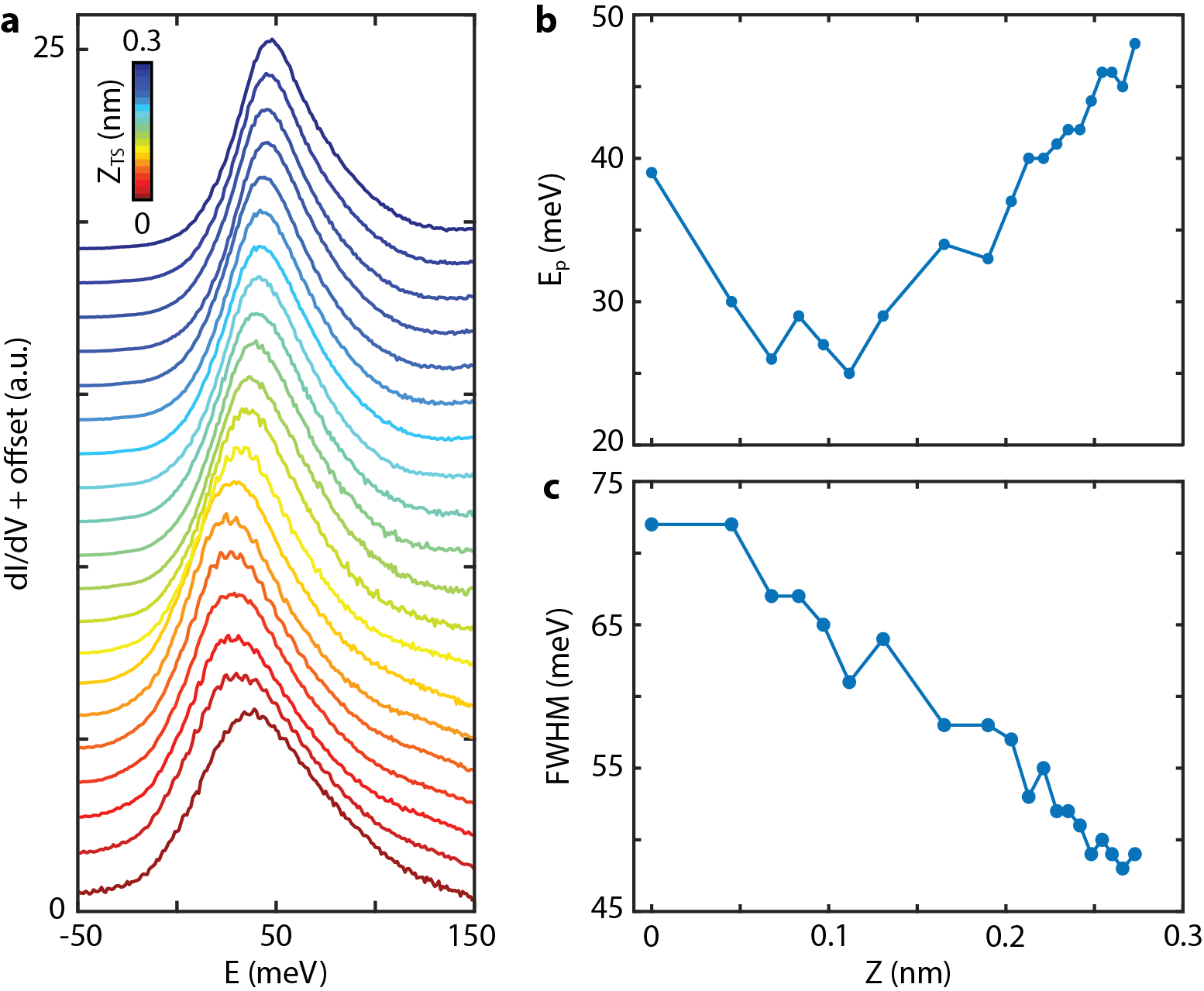}
\caption{\textbf{Non-monotonic shift of flat bands.} \textbf{a,} dI/dV profiles measured on a 'bright' CDW site (same site as in Fig.3\textbf{d}) while reducing the tip and the sample separation. The distance travelled by the tip towards the sample is denoted by $Z_{TS}$. \textbf{b-c,} The evolution of the quasi-particle peak position ($E_p$) and the width (FWHM) are shown in \textbf{b} and \textbf{c}, respectively. The peak position shifts non-monotonically while the width decreases continuously. Scanning parameters: $V_{set}$ = 200 mV, $I_{set}$ = 100 pA to 3900 pA in steps of 200 pA, $V_{ac}$ = 5 mV, f = 433 Hz. a.u., arbitrary unit.}
\label{fig:IZ}
\end{figure}

A higher-energy resolution measurement of Fig.3d is shown in Fig.\ref{fig:IZ}a. We again observe the non-monotonic shift of the peak position ($E_{p}$), shown in Fig.\ref{fig:IZ}b and the gradual reduction of the peak width (FWHM), shown in Fig.\ref{fig:IZ}c. Tip-induced band bending usually shifts the band monotonically, and therefore, the non-monotonic shift of the peak remains puzzling. The gradual reduction of the peak width may be interpreted as follows. Although the lower Hubbard band is nearly empty due to charge transfer from the 1T layer to the 1H layer, a part of it may still be occupied. Thus, the quasi-particle peak we observe at $\sim 70$ meV consists of both the upper and the partially occupied lower Hubbard band whose energy spitting is not resolved due to large peak widths. On bringing the tip closer to the sample, the interlayer coupling could increase, leading to further depletion of the lower Hubbard band. Therefore, fully emptying the lower Hubbard band would make the Hubbard bands degenerate, thus reducing the peak width. This proposal also explains merging the two peaks beyond the quantum phase transition in Fig.3c.

\FloatBarrier


\section{Set-point effect}
\begin{figure}[ht]
\centering
\includegraphics[scale=1]{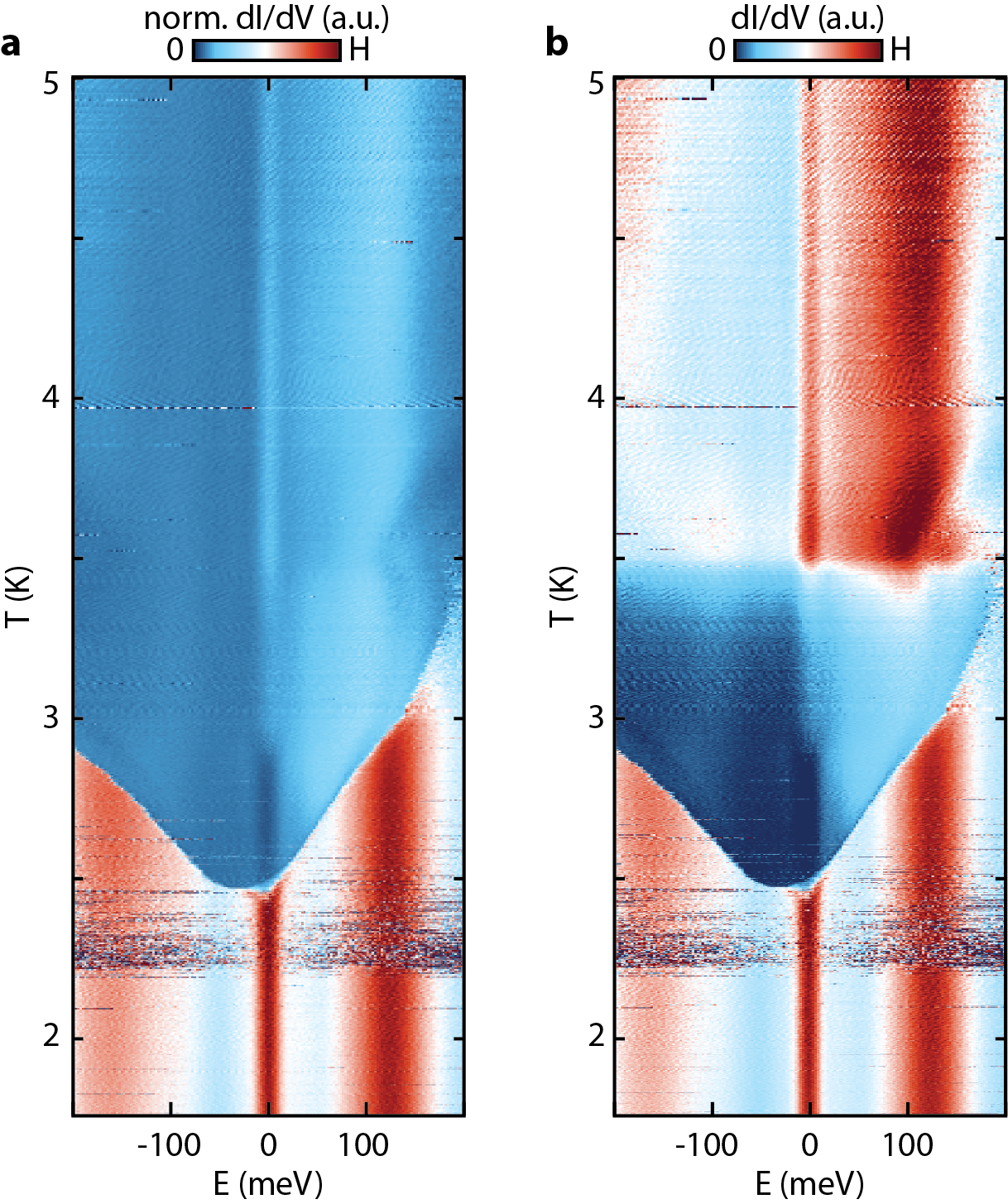}
\caption{\textbf{Temperature and Bias voltage evolution of $dI/dV$ profile with (a) and without (b) normalization.}}
\label{fig:vtraw}
\end{figure}

The raw data of the temperature sweep shown in Fig.\ref{fig4}a (also Fig.\ref{fig:vtraw}a for comparison) is shown in Fig.\ref{fig:vtraw}b. In it we find a sharp change in dI/dV that happens at 3.5 K across all bias. It is an intriguing artifact of the STM measurement (Fig.\ref{fig:vtraw}) in which the tip repositions its height in response to a change in the electronic spectrum. The feedback loop fixes the tip height to maintain a desired tunneling current under a fixed bias voltage, here set to 200 meV. That tunneling current is affected mainly by the tunneling matrix element, which is exponentially dependent to distance between the tip and the sample, as well as the integrated DOS from the Fermi level up to the bias voltage above it. If for any reason the sample DOS changes, so will the that integral and the feedback loop will response by readjusting the tip height to maintain constant current. What happens at 3.5 K is that the transition line crosses the parking bias of 200 meV. Once that happens the electronic DOS in the sample undergoes the phase transition from Kondo state at lower temperatures to DFB state above. the non-dispersive step at 3.5K results from repositioning the tip height to maintain constant current for the changing electronic phase. It is a unique manifestation of the well known set-point effect that occurs because the tip is part of the quantum electronic system it probes. However, this is a measurement artifact, and therefore we normalize it out in Fig.\ref{fig4}a by fixing a constant dI/dV value at a certain energy (-20 meV) within the DFB high-temperature state (ie 2.5K and above).

\FloatBarrier

\section{Thermodynamics of the first-order transition} \label{smthermo}
To model a first-order phase transition as function of temperature $T$ and voltage $V$, we use a Taylor expansion of the free energy of the two phases, $i=1,2$, as function of temperature $T$ and applied voltage $U$.
\begin{align}
    F^i(T,U)\approx F^i_0 -(T-T_0) S^i_0-Q^i U-\frac{1}{2} C^i U^2
\end{align}
where $S^i$ is the entropy of phase $i$, $C_i$ encodes the capacitance of the system in the presence of the tip and $Q_i$ with the units of charge denotes a shift of energy linear in the voltage, discussed in more detail below. Note that the sign of the capacitative correction is negative as the system gains energy by moving charge between system and tip. We assume that $S^2>S^1$, thus one obtains a transition from phase $1$ to phase $2$ upon increasing the temperature.

The thermodynamic first-order transition occurs at a voltage-dependent transition temperature $T_c$ obtained from 
\begin{align}
    F^1(T_c,U)= F^2(T_c,U).
\end{align}
Chosing $T_0$ such that $F^1(T_0,0)=F^2(T_0,0)$, we get
\begin{align}
   T_c \approx T_0 + \frac{1}{\Delta S}\left(-\frac{ \Delta C }{2}  U^2-\Delta Q\, U\right)
\end{align}
with  $\Delta S=S_2-S_1>0$, $\Delta C=C_2-C_1$, $\Delta Q=Q_2-Q_1$. From a fit to the experiment we find
\begin{align}
    \frac{\Delta C}{\Delta S}\approx - 60 \frac{\text{K}}{\text{V}^2}, \qquad   \frac{\Delta Q}{\Delta S} \approx -1.7  \frac{\text{K}}{\text{V}}\approx -1.5\cdot 10^{-4}\, \frac{|e|}{k_B}.\label{numbers}
\end{align}
where we used in the last line microscopic units, the electron charge $e$ and the Boltzmann constant $k_B$, resulting in a small prefactor. This indicates either an almost complete compensation of the change of polarization arising from the transfer of charge beween the layers or a very large entropy of the high-T phase. Note that the phase diagram was measured by reducing the voltage starting from the parking bias $V_b=200\,$mV. Therefore we cannot rule out that a possible hysteresis may affect the value (and even the sign) of $\frac{\Delta Q}{\Delta S}$ but we can exclude a substantially larger value of this quantity.


We can obtain a rough estimate of $\Delta Q$ by using that the electric field at the surface of the sample is of the order of $U/d$ for a flat tip or $2 U r_0/d^2$ for a very sharp tip with radius $r_0$, where $d\approx 1$\,nm is the distance of the tip from the surface. If the polarization of the two phases differs by $\Delta P$, then the field-induced difference of energies is given by the product of field and polarization. If we assume, for example, that $\Delta P$ arises from the shift of an electron charge by $1$\,\AA, we obtain $\Delta Q/e \sim 0.001\dots0.1$. Comparing this with Eq.~\eqref{numbers}, we conclude that either the difference of polarization is much smaller or the entropy difference of the two phases is relatively large, of the order of $10^2\dots 10^3\,k_B$.

When discussing the sign of $\Delta C$, it is important to realize that the largest contribution to the capacitance is geometric in origin, $C^i\approx C_{\text{geo}}$, and fixed essentially by the shape of the tip and the underlying metalic layers. Only a small extra contribution $C^i_e$ 
arises from the local compressibility $\kappa^i_e$ of the electronic phase just below the tip. The local compressibility can be computed by evaluating to second order in $U$ the change of energy arising from a voltage-induced potential (after taking the screening by the metallic substrate into account).
As $1/C^i=1/C_{\text{geo}}+\kappa^i_e$, we conclude that 
\begin{align}
\Delta C \approx -C_{\text{geo}}^2 \left(\kappa^2_e-\kappa^1_e\right)
\end{align}
As the Kondo phase has a much higher local density of states at the Fermi energy, it costs less energy to locally add charges. Thus $\kappa^1_e<\kappa^2_e$, which explains why $\Delta C$ is negative.

\FloatBarrier

\printbibliography[title=Supplementary References]
\end{refsection}